\documentclass{aa}
 \usepackage{amsmath}
\usepackage{amstext}
\usepackage{amsfonts}
\usepackage{amssymb}
\usepackage{graphicx}
\usepackage[varg]{txfonts}
\usepackage{xspace}
\usepackage{aalongtable}
\usepackage{natbib}
\usepackage{multirow}
\usepackage{siunitx}
\usepackage{lscape}

\begin{document}
\title{New NIR light-curve templates for classical Cepheids}  

\author{
L.~Inno \inst{\ref{inst1}}$^,$\inst{\ref{inst2}}
 \and 
N.~Matsunaga\inst{\ref{inst3}} \and 
M.~Romaniello\inst{\ref{inst2}} \and 
G.~Bono\inst{\ref{inst1}}$^,$\inst{\ref{inst4}}
 \and 
A.~Monson\inst{\ref{inst5}}  \and 
I.~Ferraro\inst{\ref{inst4}} \and 
G.~Iannicola\inst{\ref{inst4}}  \and 
E.~Persson\inst{\ref{inst5}}  \and 
R.~Buonanno\inst{\ref{inst1}}$^,$\inst{\ref{inst6}}
 \and 
W.~Freedman\inst{\ref{inst5}}  \and 
W.~Gieren\inst{\ref{inst7}}   \and 
M.A.T.~Groenewegen\inst{\ref{inst8}}  \and 
Y.~Ita\inst{\ref{inst9}} \and 
C.D.~Laney\inst{\ref{inst10}}$^,$\inst{\ref{inst11}}
 \and 
B.~Lemasle\inst{\ref{inst12}} \and 
B.F.~Madore\inst{\ref{inst5}}  \and 
T.~ Nagayama\inst{\ref{inst13}}$^,$\inst{\ref{inst13b}} \and 
Y.~Nakada\inst{\ref{inst14}} \and 
M.~Nonino\inst{\ref{inst15}} \and 
G.~Pietrzy{\'n}ski\inst{\ref{inst7}}$^,$\inst{\ref{inst16}}
  \and 
F.~Primas\inst{\ref{inst2}} \and 
V.~Scowcroft\inst{\ref{inst5}}  \and 
I.~Soszy{\'n}ski\inst{\ref{inst16}} \and 
T.~Tanab\'{e}\inst{\ref{inst14}} \and 
A.~Udalski\inst{\ref{inst16}}  
}

\institute{Dipartimento di Fisica, Universit\`a di Roma Tor Vergata, 
via della Ricerca Scientifica 1, I-00133 Rome, Italy; laura.inno@roma2.infn.it \label{inst1} \and
European Southern Observatory, Karl-Schwarzschild-Str. 2,D-85748 Garching bei Munchen, Germany \label{inst2} \and
Department of Astronomy, School of Science, The University of Tokyo,7-3-1 Hongo, Bunkyo-ku, Tokyo 113-0033, Japan \label{inst3} \and
INAF--OAR, via Frascati 33, I-00040 Monte Porzio Catone, Rome, Italy \label{inst4} \and
Observatories of the Carnegie Institution of Washington 813 Santa Barbara Street, Pasadena, CA 91101\label{inst5} \and
INAF-Osservatorio Astronomico di Collurania, via M. Maggini,I-64100 Teramo, Italy \label{inst6} \and
Departamento de Astronomia, Universidad de Concepci\`{o}n, Casilla 160-C, Concepci\`{o}n, Chile \label{inst7} \and
Koninklijke Sterrenwacht van van Belgi\"e, Ringlaan 3, B-1180, Brussel, Belgium\label{inst8}\and
Astronomical Institute, Graduate School of Science, Tohoku University,   6-3 Aramaki, Aoba-ku, Sendai, Miyagi 980-8578, Japan\label{inst9}\and
South African Astronomical Observatory, P.O. Box 9, Observatory 7935, South Africa\label{inst10}\and
Department of Physics and Astronomy, N283 ESC, Brigham Young University, Provo, UT 84601, USA \label{inst11}\and
Astronomical Institute Anton Pannekoek, Science Park 904, PO Box 94249, 1090 GE, Amsterdam, The Netherlands\label{inst12}\and
Department of Astrophysics, Nagoya University, Furo-cho, Chikusa-ku,
Nagoya, Aichi 464-8602\label{inst13}\and
Department of Physics and Astronomy, Graduate School of Science and
Engineering, Kagoshima University, 1-21-35 Korimoto, Kagoshima 890-0065,
Japan\label{inst13b}\and
Institute of Astronomy, School of Science, The University of Tokyo,   2-21-1 Osawa, Mitaka, Tokyo 181-0015 \label{inst14}\and
INAF-Oss. Astr. di Trieste, Via Tiepolo 11, 34131 Trieste, Italy\label{inst15}\and
Warsaw University Observatory, Al. Ujazdowskie 4, 00-478 Warszawa, Poland \label{inst16}}
\date{\centering drafted   / Received / Accepted 10/12/14  }

\titlerunning{New NIR light-curve templates for classical Cepheids}
\authorrunning{Inno et al.}

\abstract{}{We present new near infrared (NIR)  light-curve templates for fundamental 
(FU, $J$,$H$,$K_{\rm{S}}$) and first overtone (FO, $J$) classical Cepheids.
The new templates together with Period--Luminosity and Period--Wesenheit (PW)
 relations provide Cepheid distances from single-epoch observations
 with a precision only limited by the intrinsic accuracy 
of the method adopted.}
{The templates rely on a very large set of Galactic and Magellanic Clouds 
Cepheids (FU, $\sim$600; FO, $\sim$200) with well-sampled NIR (IRSF data set) 
and optical ($V$,$I$; OGLE data set) light-curves. To properly trace the change 
in the shape of the light-curve as a function of pulsation period, we split the 
sample of calibrating Cepheids into ten different period bins. The templates 
for the first time cover FO Cepheids and the short--period range of FU 
Cepheids (P$\le 5$ days). Moreover, the phase zero--point is anchored to the phase 
of the mean magnitude along the rising branch. The new approach has several advantages 
in sampling the light-curve of bump Cepheids when compared with the canonical phase 
of maximum light. We also provide new empirical estimates of the NIR--to--optical  
amplitude ratios for FU and FO Cepheids. We perform detailed analytical fits 
using seventh-order Fourier series and multi-Gaussian periodic functions. 
The latter  are characterized by fewer free parameters (nine 
vs fifteen).  }
 {The mean NIR magnitudes based on the new templates are up to 80\% more 
accurate than single--epoch NIR measurements and up to 50\%  more accurate
than the mean magnitudes based on previous NIR templates,  with typical associated
uncertainties ranging from 0.015 mag ($J$ band) to 0.019 mag ($K_{\rm{S}}$ band). 
Moreover,  we find that errors on individual distance estimates for Small Magellanic Cloud Cepheids derived from NIR PW relations are essentially reduced to the 
intrinsic scatter of the adopted relations.}
{Thus, the new templates are the ultimate tool for estimating
precise Cepheid distances from NIR single-epoch observations,
 which can be safely adopted for future interesting applications, 
including deriving the 3D structure of the Magellanic Clouds.
}


\maketitle

\section{Introduction}
 Radially pulsating variables, and in particular RR Lyrae and Classical Cepheids, play a key role 
in modern astrophysics because they are robust primary distance indicators and solid 
tracers of old (t$\sim$10--12 Gyr) and young (t$\sim$10--300 Myr) stellar populations, 
respectively.   	
The radially pulsating variables when compared with canonical stellar tracers have the key advantage 
of being easily recognized by their characteristic light-curves and to provide firm 
constraints on the metallicity gradient and the kinematics of both the thin 
disk \citep{pedicelli09, luck11, luck11a, genovali13} and the halo 
\citep{kinman12}. The above evidence applies not only to the Galaxy, 
but also to nearby resolved stellar systems \citep{minniti03}. 

The main drawback in using classical Cepheids is that they have periods ranging from 
one day to several tens of days. This means that identifying and characterizing them
is demanding from the observational point of view.
An unprecedented improvement on the number of known radially pulsating 
variables was indeed provided by microlensing experiments (MACHO, EROS, OGLE)
 as a byproduct of their large-area surveys.
In particular, the ongoing OGLE IV project became a large-scale, long-term, sky-variability survey, and will further increase the variable star identification 
 in the Galactic Bulge and in the Magellanic Clouds \citep[OGLE IV,][]{sos2012}.    
These surveys have been deeply complemented by large experiments aimed at detecting 
variable phenomena covering a significant fraction of the Southern Sky, either in the 
optical, such as the ASAS Survey \citep{pojmanski02}, the QUEST Survey \citep{vivas04}, the NSVS survey \citep{kinemuchi06}, the LONEOS Survey \citep{miceli08}, the Catalina Real-time Transient survey \citep{drake09}, the SEKBO survey \citep{akhter12} and the LINEAR Survey \citep{palaversa13};  or in the near-infrared (NIR), such as the IRSF \citep{ita04}, the VVV \citep{minniti10}; or in the mid-infrared (MIR) such as the Carnegie RRL Program \citep[CRRLP;][]{freedman12}.

The intrinsic feature of current surveys is that both the identification and the 
characterization is performed in optical bands, since the pulsation amplitude is 
typically larger in the B--band than in the NIR bands. However, recent theoretical \citep{bono10} and empirical 
\citep{storm11a,storm11b,inno13,groe13} evidence indicates that NIR and MIR photometry has several indisputable advantages when compared with optical photometry:
a) it is minimally affected by metallicity dependence \citep{bono10,madore10}; 
b) it is minimally affected by reddening uncertainties;  and 
c) the luminosity amplitude is a factor of 3--5 smaller than in optical bands. 
This means that NIR observations are not very efficient in identify 
new variables, but they play a crucial role in heavily reddened regions 
\citep{matsunaga11,matsunaga13}. 
Moreover, accurate mean NIR and MIR magnitudes  
can be provided even with a limited number of phase points,  because of their 
reduced luminosity amplitudes in this wavelength regime. However, 
NIR and MIR ground--based observations are even more time-consuming than 
optical observations, because of sky subtraction. This is the reason why during the past 20 years NIR light-curve templates have been developed for RR Lyrae 
\citep{jones96} and classical Cepheids \citep{soszynski05,pejcha12}. 
The key advantage of this approach is that variables for which 
the pulsation period, the epoch of maximum, and the $B$- and/or the $V$--band 
amplitude are available, a single-epoch NIR measurement is enough to provide accurate 
estimates of their mean NIR magnitudes, which can then be used to compute their distances.
  
The most recent Cepheid NIR light-curve template was published by 
\citet[][hereinafter S05]{soszynski05}. They used 30 Galactic and 
31 Large Magellanic Cloud (LMC) calibrating Cepheids and provided 
analytical Fourier fits in the $J$, $H$, and $K$ bands by using 
two period bins for  Galactic (0.5$< \log P \le$ 1.3 and $\log P > 1.3$) 
and LMC (0.5$<\log P \le 1.1$ and $\log P > 1.1$) Cepheids.

We derived new sets of NIR light-curve templates for classical Cepheids 
covering the entire period range (0.0 $< \log P \le$ 1.8).
The advantages of the current approach compared with previous 
NIR templates available in the literature are the following: 

a) Statistics -- We collected optical and NIR accurate photometry 
for more than 180 Galactic and 500 Magellanic Cloud Cepheids. 
Among these data, we selected the light curves characterized 
by full phase coverage and high photometric quality in the 
$V$,$J$,$H$ and $K_{\rm{S}}$ bands. We ended up with a sample 
of more than 200 calibrating Cepheids.
This sample is a three times larger than the sample 
adopted by S05.   
The sample size enabled us to split the calibrating Cepheids into 
ten period bins ranging from one day to approximately 
100 days; 

b) Hertzsprung progression  -- The sample size allowed us to properly 
trace the change in light-curve morphology across the Hertzsprung 
Progression (HP). Cepheids in the period range 6$<$ P$<$16 days show
a bump along the light-curves. The HP indicates the relation between
this secondary feature and the pulsation period:
the bump crosses the light-curve from the decreasing branch to the maximum
for periods close to the center of the HP and moves to earlier phases 
for longer periods.  To properly trace the change in the shape of the light 
curve, we adopted a new anchor for the phase zero--point. The classical 
approach was to use the phase of maximum light of optical light 
curves to phase the NIR light-curves. The use of the phase of maximum light 
as zero--point ($\phi$=0) was justified by the fact that the photometry was more 
accurate along the brighter pulsation phases. 
However, this anchor has an intrinsic limit in dealing with bump Cepheids. 
At the center of the HP the optical light-curves are 
either flat topped or show a double peak. This means that from an empirical 
point of view it is quite difficult to identify the phase of maximum light. 
Moreover, the center of the HP is metallicity dependent (see Sect.~2.1). 
To overcome this problem, we decided to use the phase of the mean magnitude    
along the rising branch. This phase zero--point can be easily estimated 
even if the light-curve is not uniformly sampled;

c) First overtones -- We derived for the first time the template for 
first overtone Cepheids in the $J$ band;

d) Analytical fit -- Together with the classical analytical fits based 
on Fourier series we also provide a new analytical fit based on periodic 
Gaussian functions. The key advantage in using the latter functions is 
that the precision is quite similar to the canonical fit, but the number of 
parameters decreases from fifteen to nine.

The structure of the paper is the following:
In Sect.~2 we discuss in detail the different samples of calibrating 
light curves we adopted to derive the template for fundamental (FU) 
and first overtone (FO) Cepheids.  In particular, in Sect.~2.1 we describe 
the new technique adopted for  phasing and merge the light-curves. 
The preliminary analysis of the calibrating NIR light-curves and
 the development of the template is described in Sect.~3, while the 
analytical formula are given in Sect.~4. In Sect.~5, we discuss in 
detail the NIR--to--optical amplitude ratios that we adopted to apply 
the new templates. Sect.~6 describes the application of the templates and 
with the error budget associated with the new templates. Finally, in Sect.~7 
we summarize the results of this investigation and briefly outline possible 
future developments. 

\section{Optical and NIR data sets for calibrating Cepheids}
Our analysis is based on the largest available sample of fundamental--mode Cepheids 
with well-covered light-curves in the NIR and in optical ($V$,$I$) bands. 
This sample covers a very broad period range (1--100 days). 
We collected $J$, $H$, and $K_{\rm{S}}$band observations from four different data sets: 
\citet[][51]{laney92}  and \citet[][131]{monson11} for 
Galactic Cepheids, \citet[][92]{persson04} for LMC Cepheids,
and the IRSF survey catalog for $\sim$500 Small Magellanic Cloud (SMC) Cepheids.
\\
The optical ($V$,$I$) light-curves for the galactic Cepheids were collected from the literature
\citep[][and references therein]{laney92, berdnikov04}\footnote{see also \url{www.astronet.ru/db/varstars} and \url{www.astro.utoronto.ca/DDO/research/cepheids}}, while for Magellanic Cloud Cepheids we adopted the data from the OGLE~III Catalog of Variable Stars \citep{sos2008,sos2010}\footnote{\url{www.ogledb.astrouw.edu.pl}}.
When compared with the other microlensing surveys mentioned in the previous section, the OGLE~III Catalog provides very accurate $V$-- and $I$--band light-curves, with the typical photometric error associated with individual measurements $\le$0.008 mag and $\le$ 0.006 mag for $I$ and $V$ bands, respectively. Moreover, the sky coverage of the OGLE~III survey fully matches the coverage by the IRSF survey.\\

\noindent{\it i) Calibrating Cepheids in the IRSF/OGLE Sample (SMC)} \\
\indent During the past few years, the IRSF Survey (Ita et al., 2014, in prep.)  collected more than 
$\sim$500 complete NIR light-curves (571 $J$, 434 $H$, 219 $K_{\rm{S}}$) 
for SMC Cepheids. These Cepheids have optical ($V$,$I$) mean magnitudes, 
periods, amplitudes, and positions from the OGLE~III Catalog of Variable Stars 
\citep{sos2010}.
Typically we have more than 1,000 measurements in the 
optical and at least 100 in NIR bands for SMC Cepheids. 
This means that phasing optical and NIR light-curves is relatively simple.
The photometric accuracy of the data in the IRSF catalog is $\pm$0.02 mag 
for the brightest ($J \approx  13$) and  $\pm$0.06 mag for the faintest 
($J \approx 17$) Cepheids. To improve the quality of the calibrating 
sample, we performed a selection based on the root mean square ($rms$) 
between the individual data points and the analytical fit of the 
individual light-curves. We adopted a seventh-order Fourier series 
and the selection criterion $rms \le \frac{1}{20} A_J$,
where $A_J$ is the pulsation amplitude in the $J$ band 
($A_J$=$J_{max}-J_{min}$). The $rms/A_J$ ratio is an indication of how strongly the photometric errors of the individual observations affect the shape of the light-curve.
The threshold we chose allows us to select the most accurate light-curve while keeping a statistically
 significant sample for each bin. 
However, small changes of the adopted values do not significantly affect our results.
The selection criterion was relaxed to 
$rms \le \frac{1}{15} A_{H,K}$ for the $H$-- and $K_{\rm{S}}$--band light-curves, 
because the amplitude decreases with increasing wavelength. For Cepheids with shortest period Cepheids (1--3 days) we selected the light-curves with 
$rms \le \frac{1}{10} A_{J,H,K}$. The data for these fainter Cepheids are 
characterized  by larger photometric errors. 
The \textit{J,H,K$_{\rm S}$} measurements were transformed into the 2MASS NIR 
photometric system following \citet{kato07}. 
However, the corrections adopted for the transformations between different 
NIR photometric system transformations are smaller than a few hundredths of 
magnitude and affect neither the shape nor the amplitude of the light 
curves. \\ 

\begin{figure}[!ht]
\begin{center}

\includegraphics[width=0.90\columnwidth]{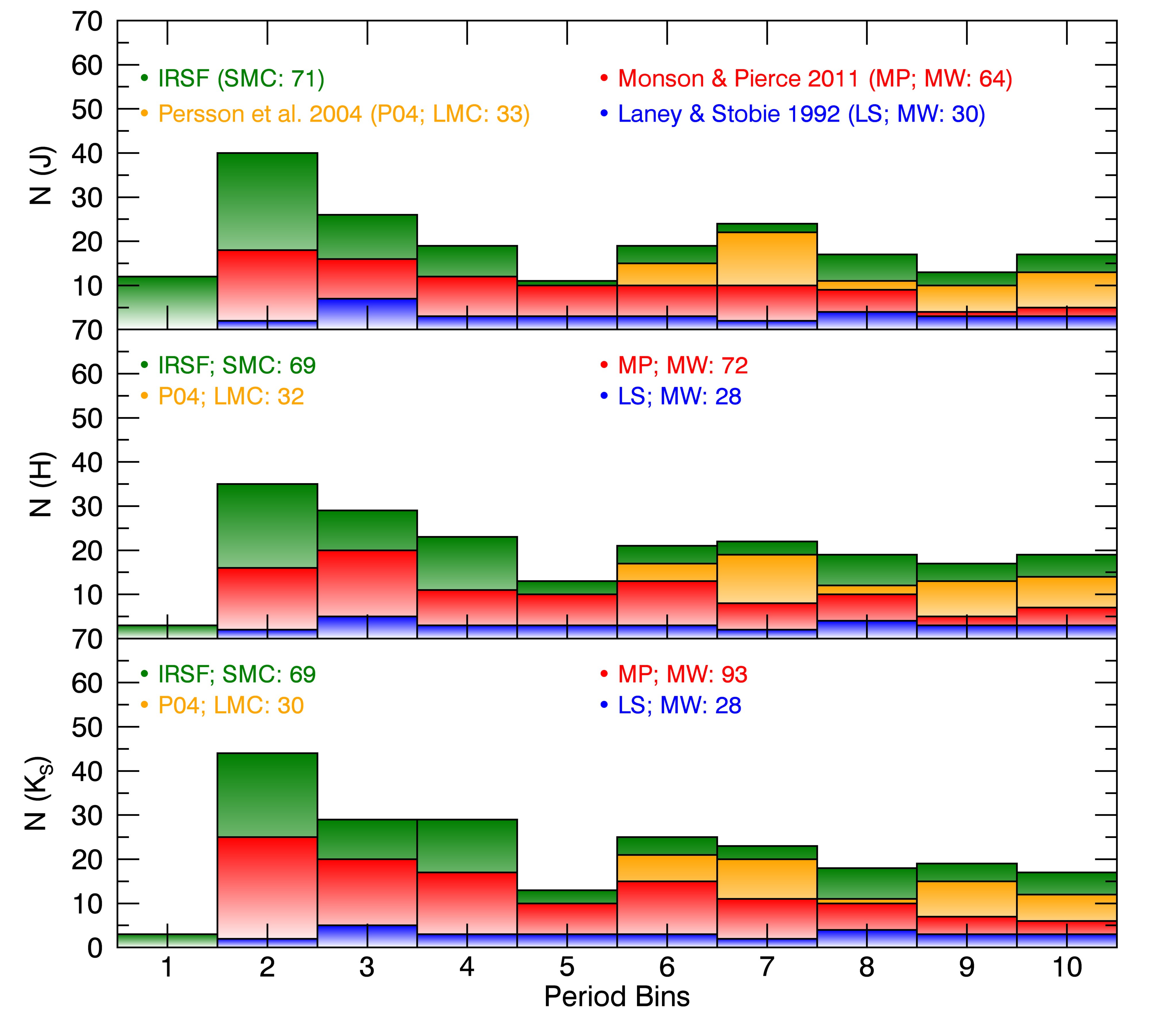}
\caption{Period distribution for the calibrating Cepheids in our sample for the 
three different bands $J$ (top), $H$ (middle), and $K_{\rm{S}}$ (bottom). 
The color coding indicates different data sets (red: MP sample; 
blue: LS sample; green: SMC sample; orange: P04 sample). The total number of 
calibrating Cepheids per data set are also labeled. See text for more details.}
\label{f1}
\end{center}
\end{figure}

\noindent{\it ii) Calibrating Cepheids in the Persson et al. 2004 Sample (P04)} \\
\indent This sample is based on the $J$,$H$, and $K$ light-curves 
for 92 Cepheids in the LMC published by \citet[][P04]{persson04} 
and the  $V$ and $I$ photometric data available in the OGLE Catalog for 60 of them. 
The photometric precision of the data in the P04 catalog is $\pm$0.02 mag 
for the brightest ($J \approx  12$) and  $\pm$0.06 mag for the faintest 
($J \approx 14$) Cepheids. 
We included Cepheids with complete coverage 
of the light-curve in the three bands (more than 20 phase points) and with low $rms$
(i.e. $\le$ 0.4 mag in the J band), for a total of $\sim$30 selected Cepheids in the 
P04 sample.  These Cepheids have periods between 6 and 50 days, 
thus increasing the number of calibrating Cepheids ($\gtrsim$50\%) in the long-period regime (see Fig.~\ref{f1}). To transform the NIR measurements from the 
original LCO photometric system into the 2MASS photometric system,
we adopted the relations given by \citet{carpenter01}.\\

\noindent{\it iii) Calibrating Cepheids in the Laney \& Stobie Sample (LS)} \\
\indent This sample includes 51 Galactic Cepheids with optical and NIR 
light curves with periods ranging from 3 to 69 days 
\citep{laney92,laney94}. This is the most accurate photometric 
sample for Classical Cepheids available in literature, with intrinsic 
errors ranging from about $\pm$0.004 mag for the brightest 
($J \approx 3$) to $\pm$0.011 mag for the faintest ($J \approx  8$). 
The optical light curves are also highly accurate,
with typical values of the $rms \le \frac{1}{100} A_{V}$.
However, we still performed a selection based on the number of available 
phase points (we need at least 15 measurements for the Fourier fit) 
in the optical and the NIR light-curves, for a total of $\sim$30 
selected Cepheids in the LS sample.
The \textit{J,H,K$_{\rm S}$} measurements in \citet{laney92} 
are in the SAAO photometric system.
We have converted them into the 2MASS photometric system 
by applying the transformation equations given by  \citet{carpenter01}.
\\

\noindent{\it iv) Calibrating Cepheids in the Monson \& Pierce Sample (MP)} \\
\indent This sample is based on NIR photometric measurements for 
131 Northern Galactic Cepheids \citep{monson11}. 
These Cepheid light-curves are sampled with an average of 22 measurements 
per star with an associated photometric error of $\pm$0.015 mag. 
However, we selected only the light curves with more than 17 measurements in each
NIR band, for a total of 64, 72, and 93 light-curves in the $J$,$H$, and $K_{\rm{S}}$ bands, respectively. 
The original \textit{J,H,K$_{\rm S}$} data were taken in the BIRCAM system
and transformed into the 2MASS photometric system
by applying the equations given by \citet{monson11}. 
For all the Cepheids in this sample, $V$ and $I$ light curves were collected 
from the literature \citep[][and references therein]{berdnikov04}. 
We did not perform any selection on the optical light-curves, 
because no high accuracy in the $V$-band data is required by our method.
However, the $rms$ is  always better than $\frac{1}{15} A_{V}$ for all the Cepheids in
this sample.
\\

In total, we collected a sample of light-curves that includes more 
than 200 calibrating Cepheids and is three times larger 
than the sample adopted by S05 (60 Cepheids) to derive the NIR light-curve 
templates for classical Cepheids. 

To further improve the sampling of the light curve over the entire period 
range and to reduce the $rms$ of the light-curve templates, the sample of 
Galactic and Magellanic calibrating Cepheids was split into ten period 
bins. 

Note that the approach we adopted is completely reddening independent. 
In particular, the period is the safest diagnostic to bin the calibrating sample, because it
can be easily measured with high accuracy, it does not depend on the wavelength, and it is not affected by reddening.
This means that the binning in period will not introduce any systematic effect
when combining optical and NIR photometric data from different instruments.  
Moreover, theoretical predictions  \citep{marconi05} clearly show that the light-curve shape changes 
with the mass at fixed chemical composition and luminosity and that the period is the best observable to account for this trend.

The adopted period ranges and the number of calibrating Cepheids 
per bin are listed in Table~\ref{tab_bins}. 

\begin{table} 
\caption{Adopted period bins.\label{tab_bins}}	
\centering 
\begin{tabular}{clccc}
\hline\hline	
Bin& Period range [days]& N$_J$ & N$_H$& N$_{Ks}$ \\
\hline
1 & 1--3 &12&3&3\\
2 & 3--5 &40&35&44\\
3 & 5--7 &26&29&29\\
4 & 7--9.5 &18&23&29\\
5 & 9.5--10.5 &11&13&13\\
6 & 10.5--13.5 &19&21&25\\
7 & 13.5--15.5 &24&22&23\\
8 & 15.5--20 &17&19&18\\
9 & 20--30 &14&17&19\\
10 & 30--100 &16&19&17\\
\hline
   & 1--100 &197&201&220\\
   \hline 
\end{tabular} 
\end{table}

There are typically twenty Cepheids per period bin 
with two exceptions:  bin~1 (P$\le$3 days), for which 
we have fewer than a dozen objects, and  bin~5 (9.5$\ge$ P $<$10), 
for which the number of Cepheids ranges from 11 ($J$--band) to 
13 ($K_{\rm{S}}$ band). 
Fig.~\ref{f1} shows the histograms of the calibrating Cepheids 
in $J$ (top), $H$ (middle) and in the $K_{\rm{S}}$ (bottom) band. 
Cepheids belonging to different data sets are plotted with 
different colors.

A similar selection was also performed for FO Cepheids.
The IRSF monitoring survey collected $\sim$231 complete NIR light-curves for 
FO Cepheids (231 $J$, 85 $H$, 10 $K_{\rm{S}}$) with periods 
ranging from 0.8 to 4 days. We selected from those the calibrating light curves  
by adopting the following criterion: $rms_J \le \frac{1}{10} A_J$.
Again, the threshold was chosen to guarantee the good photometric quality 
of the calibrating light curves.  
However, because of the limited photometric accuracy of individual measurements 
 compared with FU light-curves, the final sample of calibrating 
FO Cepheids only includes ten $J$-band light curves, with periods ranging from 1.4 to 3.5 days. 
We did not apply any binning in period for FO Cepheids, because the 
shape of the light-curve in this period range is almost exactly sinusoidal.

\subsection{Phasing the light-curves}

Precise period determinations are required to derive correct phase shifts 
between optical and NIR light curves. The constraint is less severe if optical 
and NIR time-series data are collected in the same time interval.  
The $V$ and $I$ photometric data for Galactic Cepheids cover a time interval that 
ranges from several years to more than 20 years. Thus, we adopted the new period 
estimate published by \citet[][G13]{groe13} for all the Cepheids 
in the LS sample. The G13 sample includes $\sim$130 Galactic Cepheids, and  
50\% of them are in common with the MP sample. 
The light curves from MP were phased by adopting the 
period given in the General Catalog of Variable Stars (GCVS). To check the consistency 
of the period listed in GCVS, we compared them with periods estimated by using either 
the Lomb-Scargle algorithm \citep{press89} or the PDM2 \citep{stell11}. The difference 
between the two sets of periods are about of $10^{-3}$ days, thus we 
adopted the GCVS for all the MP Cepheids not included in the G13 catalog. 

The phase of the light-curve is usually defined by
\begin{equation}
\phi^{V}_{obs} = \bmod \left( \frac{JD^{V}_{obs}- JD^V_{max}}{P} \right),
\end{equation}
where $JD^{V}_{obs}$ is the epoch of the observation and 
$JD^V_{max}$ is the epoch of the maximum in the $V$ band. 
The epoch of maximum in the $V$ and $I$ bands for the 
LMC and SMC Cepheids is available from the OGLE~III catalog, 
while for the Galactic Cepheids in LS and the 50\% in MP 
we used the epoch of the maximum estimated by G13. 
To estimate the maximum brightness for the Cepheids 
for which the epoch of maximum was not available, 
we fitted the V--band light-curves with a seventh-order Fourier series.
\begin{figure}[!ht]
\begin{center}
\includegraphics[width=0.90\columnwidth] {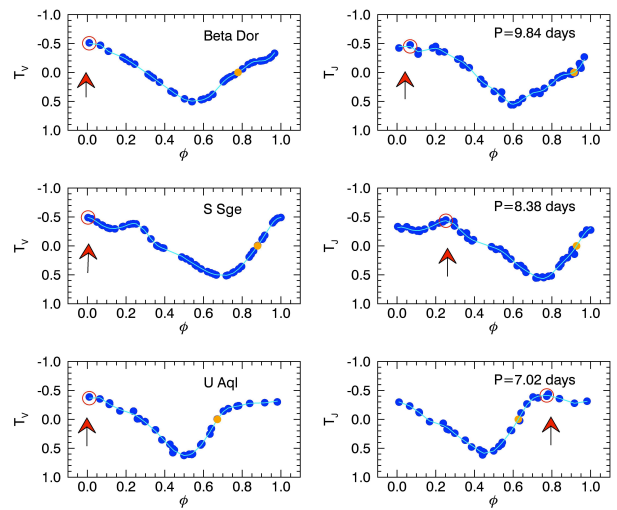}
\caption{$V$--band (left panels) and $J$--band (right panels) 
normalized light-curves for Galactic bump Cepheids in the LS sample. The 
period increases from  bottom to top. For all these light-curves 
the phase $\phi$=0 was fixed according to the maximum brightness
in the $V$ band and was marked with the arrow and the red circle.
However, the $J$--band maximum brightness, marked by the arrow, 
moves across the light-curves, and occurs at later phases than
the optical maximum. The drift in phase arises because the secondary bump can be brighter than the true 
maximum, that corresponds to the phase of minimum radius along 
the pulsation cycle. The overplotted orange dots show the 
position of the mean magnitude along the rising branch, which 
we adopt as the new phase zero--point ($\phi$=0).}
\label{f2}
\end{center}
\end{figure}

The identification of the epoch of maximum in our optical light-curves 
is straightforward, because of the very accurate time sampling of $V$- and 
$I$-band light-curves of calibrating Cepheids. However, light-curves of 
extragalactic Cepheids are covered with a limited number of phase
points, typically fewer than two dozen \citep{sandage06,bono10,madore10}.
The problem becomes even more severe for bump Cepheids, because 
 the bump moves from the decreasing to the rising 
branch in a narrow period range. 
As mentioned above, the light-curve at the center of HP 
becomes either flat topped or double peaked with the real maximum and 
the bump located at close phases. 
Fig.~\ref{f2} shows the normalized optical ($V$; left) and NIR 
($J$; right) light-curves for three Galactic Cepheids located across the 
center of the HP and with periods ranging from 7.02 (U Aql, bottom) 
to 9.84 days($\beta$ Dor, top) . If we apply the strict definition of 
epoch of maximum the phase shift between the optical and the NIR light-curve
ranges from $\sim$0.1 ($\beta$ Dor) to $\sim$0.8 (U Aql). The red circles 
and the red arrows show that the identification of the luminosity maximum 
is hampered by photometric errors and by the fact that the bump can be 
brighter than the true maximum that corresponds to the phases of maximum 
contraction (minimum radius).

The scenario is further complicated because theory 
\citep{bono00,marconi05} and observations 
\citep{moskalik92,welch97,beaulieu98,moskalik00} indicate that the 
center of the HP is anticorrelated with the metal content. It moves 
from 9.5 days for Galactic Cepheids to 10.5 and to 11.0 days for LMC 
and SMC Cepheids.   
For a more quantitative analysis of the physical 
mechanism(s) driving the HP refer to 
\citet{bono02,marconi05} and  refer to \citet{sos2008,sos2010} for a thorough 
analysis of the observed light-curves.      

To overcome this problem, we decided to use a different zero--point 
to phase optical and NIR light-curves. Our phase zero--point
is defined as the phase of the mean magnitude along the rising 
branch of the $V$--band light-curve

\begin{equation}
\phi_{obs}^V = \bmod \left( \frac{JD^V_{obs}- JD^V_{mean}}{P}\right),  
\end{equation}

\begin{figure}[!h]
\begin{center}
\includegraphics[width=0.90\columnwidth] {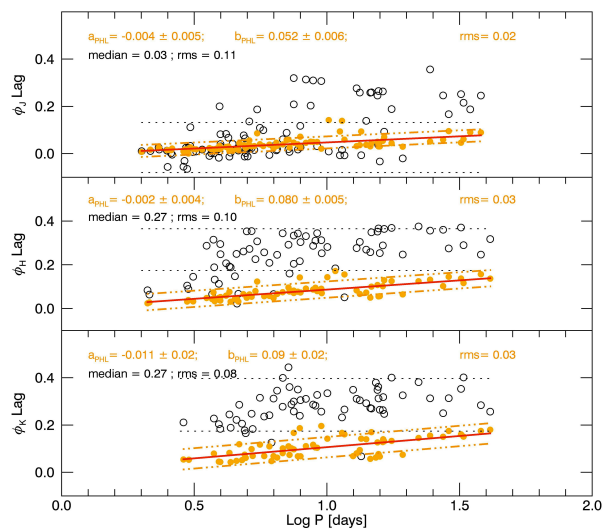}
\caption{Top: Phase lag between the $V$-- and the $J$--band light-curve of 
SMC Cepheids. The black open circles are estimated by adopting the epoch of maximum brightness as phase zero--point, 
while the orange dots by adopting the new phasing, i.e. the epoch 
of the mean--magnitude along the rising branch.  The red lines show the linear fit  
of the orange data set. The $rms$ (dashed lines) are also overplotted.
The values of the medians and of the $rms$ are labeled in the top of the panel. 
Middle: Same as the top, but for the $H$ band. 
Bottom:  Same as the top, but for the $K_{\rm{S}}$ band.}
\label{f3}
\end{center}
\end{figure}

We selected this phase point, because the mean magnitudes are more precise 
than the maximum brightness in Cepheids with modest phase coverage.  
The new phase zero--point allows us to highly improve the 
precision of the light-curve template in the period bins located across 
the bump (bin~4 to 6). A more detailed discussion of  the impact 
of the new phase zero--point is given in Sect. 3.1.
The top panel of Fig.~\ref{f3} shows the comparison between the phase lags of $V$ and $J$ light 
curves for a sample of SMC Cepheids based on the epoch of the maximum 
(Eq.~1, black open circles) and on the epoch of the mean (Eq.~2, orange dots). 
The same comparison for the $H$ and $K_{\rm{S}}$--bands are shown 
in the middle and bottom panels of the same figure.
Data plotted in this figure clearly show the advantages in using the 
epoch of the mean magnitude as the phase zero--point. 
%

i) The phase lag anchored to the epoch of the mean can be approximated
by linear relations on a broad period range.  
The intercept values of the phase lag are almost zero for all the bands and it slightly 
increases from $\sim$0 for the $J$and $H$ bands to 0.011 for the $K_{\rm{S}}$ band.  
The slope also systematically increases from 0.05 for the $J$ band, 
to 0.08 for the $H$ and 0.09 for the $K_{\rm{S}}$ bands.
Moreover, the standard deviations based on the epoch of the mean magnitude are 
at least a factor of two smaller than the standard deviations of the phase lags 
based on the epoch of maximum light (0.02 vs 0.11 in the $J$ band). 
Thus, using  the epoch of the maximum introduces a spurious shift 
in the epoch of the maximum of the NIR light-curves of bump Cepheids, and in 
turn a systematic error in the estimate of their mean NIR magnitudes. 

ii)  The zero--point and the slopes of the linear relations to predict the 
phase lag between optical and NIR light-curves are 
similar for Magellanic and Galactic Cepheids.

iii) The slope  of the light-curve's rising branch is steeper than that of the decreasing branch. This means that the error on the estimate of the mean magnitudes propagates into a smaller error in the phase determination. For this reason, the epoch of the mean along the rising branch provides a more accurate phase zero--point than the phase along the decreasing branch. However, the shape of the 
light curves changes once again for periods longer than the HP. 
The rising branch of Cepheids with periods longer than 15.5 days is shallower 
than the decreasing branch. This means that the latter provides a more solid 
phase zero--point. However, this problem only affects a minor fraction of our 
sample ($\approx$10\%), therefore we adopted the phase of the mean magnitude 
along the rising branch.    
 
The phase lags between $V$- and $J$-band  FO Cepheid light-curves 
are similar to those of the FU Cepheid, with a median value of 0.05.  
\begin{figure*}[!t]
\begin{center}
\includegraphics[width=0.8\textwidth, height=0.4\textheight] {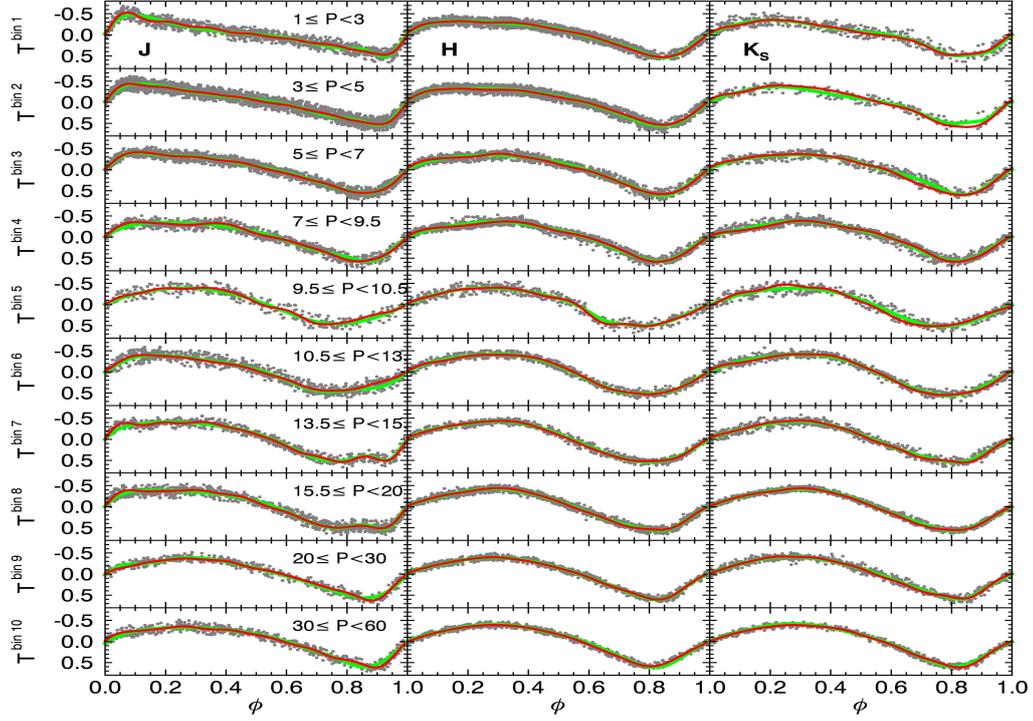}
\caption{Merged light-curves ($T^{bin}$) for the $J$ (left), $H$ (middle) and $K_{\rm{S}}$ bands
(right) for the ten period bins of FU Cepheids (see Eq.~5 and the following text for more details).  
The range of periods in days is labeled in the top right corner of the left panel for each bin.  
The G3 (green line) and the F7 (red line) templates are also shown.}
\label{f4}
\end{center}
\end{figure*}

To estimate the phase of the mean magnitude for the entire sample of calibrating 
Cepheids, we fit the V--band light-curves transformed into intensity with a 
seventh-order Fourier series. The mean in intensity was estimated as the 
constant term of the analytical fit and eventually transformed into magnitude. 
The comparison with the $V$--band mean magnitudes provided by \citet{sos2010} 
by using the same definition indicates that the difference is at most of the 
order of a few hundredths of magnitude.      
The luminosity minimum and the luminosity maximum were estimated as the mean 
of the three closest observed point located across the luminosity maximum and 
the luminosity minimum of the analytical fit. We adopted this approach because  
the analytical fit in the period ranges in which the light-curves show secondary 
features (3 $\le$ P $\le$ 5 days;  7 $\le$ P $\le$ 15 days) slightly 
underestimate the luminosity amplitude. Moreover, the error associated with the 
 amplitude estimated by adopting this approach does not depend on the analytical fit, 
but is given by the propagation of the photometric error of the observations. 
 
After estimating the epoch of the mean V-band magnitude ($\phi_{mean}^V$) for  
FU and FO calibrating Cepheids, the epoch of the mean 
magnitude in the NIR bands is given by

\begin{equation}
JD_{mean}^{J,H,Ks}=JD_{mean}^{V}+ P \times \phi_{Lag}^{J,H,Ks},
\end{equation}
where $\phi_{Lag}^{J,H,Ks}$ is a constant for the different bands. Its 
value for FU Cepheids with periods shorter than 20 days is 0.03 ($J$), 
0.07 ($H$), and 0.13($K_{\rm{S}}$), while for longer periods it is: 
0.06 ($J$), 0.14 ($H$), and 0.16($K_{\rm{S}}$). The phase lag in the $J$ band 
for FO Cepheids is  0.05.
This equation can also be written in terms of $JD_{max}^{V}$ and $\phi_{mean}^{V}$:
$$JD_{mean}^{J,H,Ks}=JD_{max}^{V}+P \times (\phi_{mean}^{V}+\phi_{Lag}^{J,H,Ks})$$.
Thus, the pulsation phase can also be defined as

\begin{equation}
\phi_{obs}^{V,J,H,Ks} = \bmod \left( \frac{JD_{obs}^{V,J,H,Ks}-JD_{mean}^{V,J,H,Ks}}{P}\right),   
\end{equation}
where the symbols have their usual meanings. 
The name, the period, the $V$, $J$, $H$, $K_S$ mean magnitudes, the amplitude pulsations, and the 
epoch of the mean magnitude along the rising branch for the entire sample of calibrating FU Cepheids are listed in Table~\ref{tab_cat_fu}.  
The same parameters for FO calibrating Cepheids are listed in 
Table~\ref{tab_cat_fo}.  

\section{Merged NIR light-curves of calibrating Cepheids}

To compute the light-curve template for FU Cepheids in the different period bins, 
we performed a fit with a seventh-order Fourier series of the $V$,$J$,$H$,$K_{\rm{S}}$ 
light curves of the calibrating Cepheids.
The analytical fit provides several pulsation parameters: 
mean magnitude\footnote{Throughout the paper, mean magnitude means a mean in intensity 
transformed into magnitude.}, pulsation amplitude, and the phase of the mean along the 
rising branch.
The fit with a seventh-order Fourier series is the most often used for 
Classical Cepheid light-curves \citep{laney92,sos2008,sos2010}. Analytical fits 
with lower order Fourier series cause an underestimate of FU Cepheid luminosity 
amplitudes. On the other hand, higher order analytical fits cause minimal 
changes (a few thousandths of magnitude) in the luminosity amplitudes and 
in the mean magnitudes.
Following the same approach as adopted by S05, we normalized the 
light curves in such a way that the mean magnitude 
is equal to zero and the total luminosity amplitude is equal to one. 
In particular, for the $J$ band, the normalized light-curve is defined as 

\begin{equation}
T_{J,l} = (J_{i,l} - <J>_l)/A_{J,l},
\end{equation}
where $J_i$ are the individual measurements in the $J$ band,  $<J>$ is the mean 
magnitude and $A_J$ is the luminosity amplitude of the variable in the $J$ band 
for the $l$-th light-curve.   

This approach allowed us to compute the merged light-curve for each period bin 
($T_J^{bin}$). The merged light-curves for 
the ten period bins in the $J$ (left), $H$ (middle), and $K_{\rm{S}}$ (right) are shown 
in Fig.~\ref{f4}. Data plotted in this figure clearly show that current 
NIR data set properly cover the entire pulsation cycle in the three different 
bands in the short and the long period range.  Moreover, the intrinsic 
scatter at fixed pulsation phase is quite small and ranges from $\sim$0.03 to 
$\sim$0.05 over the entire data set. This evidence underlines the 
photometric precision and homogeneity of the adopted data sets together with 
the selection of calibrating Cepheids.   

\begin{figure}[!hb]
\begin{center}
\includegraphics[width=0.90\columnwidth, height=0.8\columnwidth] {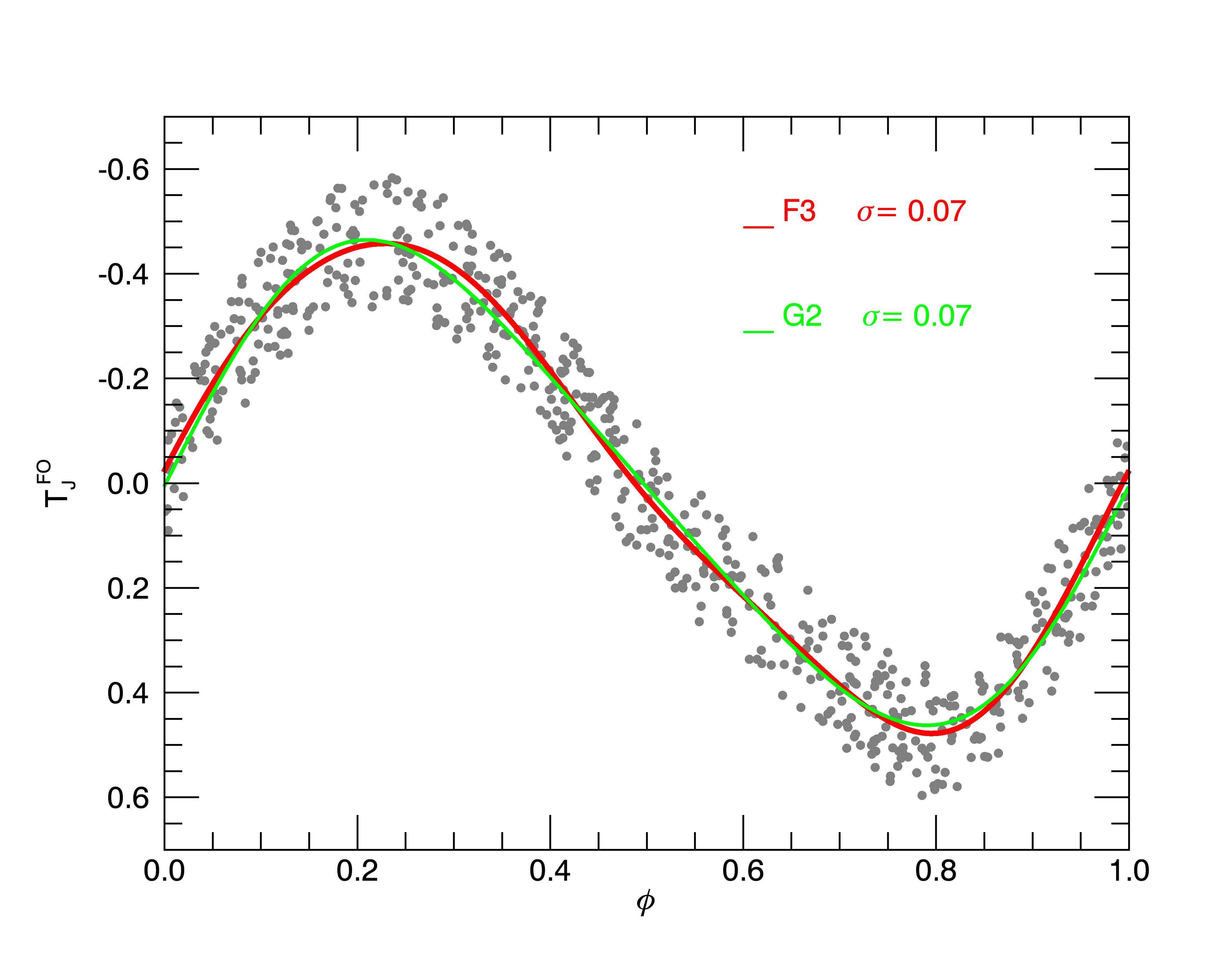}
\caption{Merged light-curve for FO Cepheids. The F3  (red line) and G2 (green line) 
templates are also shown.}
\label{f5}
\end{center}
\end{figure}

We adopted the same approach for the FO calibrating Cepheids. 
To estimate the main physical parameters, we fit the FO light-curves 
with a third-order Fourier series, because they have a 
sinusoidal shape in the adopted period range (see Fig.~\ref{f5}).

\section{Analytical fits of the merged NIR light-curves of calibrating Cepheids}

The  precision of the mean NIR magnitudes based on light-curve templates 
depends on the accuracy of the analytical fits in reproducing the shape 
of the individual light-curves in the different period bins. 
The fit of the merged light-curves was performed by using seventh-order Fourier series

\begin{equation}
{\rm F7(\phi)}= A_0 + \sum_{i=1}^7 A_i \cos (2 \pi i \phi + \Phi_i).
\end{equation}

Fitting the light curves with Fourier series is a very popular approach for 
both regular and irregular variables. They have many advantages, but also 
several limits. In particular, the F7 templates (red lines) for the bin~1 
in the $J$,$H$, and $K_{\rm{S}}$ bands --see top panels of Fig.~4-- show 
several spurious ripples along the decreasing branch of the light-curve. 
Moreover, the F7 templates for the period bins located across the bump 
display a wiggle close to the phases of maximum brightness (bin~5) and 
a stiff trend close to the phases of minimum brightness (bin~7). 

To provide an independent approach for the analytical modeling of the 
light curves, we adopted a fit with multiple periodic Gaussian functions

\begin{equation}
{\rm G3(\phi)}= \sum_{i=1}^3 G_i \exp \left[ \frac{- \sin \pi (\phi- \Gamma_i)}{\tau_i}\right]^2.
\end{equation}

\begin{figure}[!ht]
\begin{center}
\includegraphics[width=0.90\columnwidth] {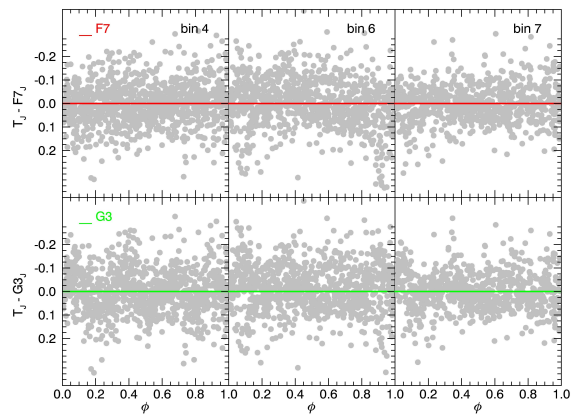}
\caption{From left to right:  residuals of the $J$--band merged light-curves 
(silver dots) obtained with the new templates: F7 (red line; top) and 
G3 (green line; bottom) for the period bin~4 (7--9.5 days), bin~6 (10.5--13.5 days) 
and bin~7 (13.5--15.5 days). 
The residuals attain vanishing mean values for the F7 and the G3 templates.}
\label{f6}
\end{center}
\end{figure}

We called these analytical functions PEGASUS, because they provide  
PEriodic GAuSsian Uniform and Smooth fits. The key advantage of these functions 
is that they follow the features of the light-curves, but the wings remain 
stiff. The fits of the calibrating light-curves with the linear combination 
of three PEGASUS functions are plotted as green lines in Fig.~\ref{f4} and are very accurate over the entire period range. 
The fits based on PEGASUS  show two indisputable advantages over the Fourier series:  a) the PEGASUS fit (G3) only requires nine parameters, 
while the Fourier fit (F7) needs 15 parameters, and b) the G3 fit does not show the ripples in either the short period bins (bins~1 and 2; 
see Fig.~\ref{f4}) or across the Hertzsprung progression (bins~4, 6, and 7; 
see Fig.~\ref{f6})

However, the standard deviations of the individual fits and the errors on the 
coefficients in G3 and in F7 fits attain similar values. The standard deviations 
are on average on the order of a few hundredths of magnitude, wile the errors 
of the coefficients are smaller than one thousandth of magnitude. This is the main reason
why we decided to provide the analytical fits for both of them. The coefficients 
$A_i$ and $\Phi_i$ for the F7 fits and the $G_i$,  $\Gamma_i$ and $\tau_i$ coefficients for the G3 templates are given in Table~\ref{tab_c}.

For the FO light-curve template, we adopted a third-order Fourier  (F3) series 
and a second-order PEGASUS function (G2) to fit the merged light-curves.
Fig.~\ref{f5} shows the $J$--band template for FO Cepheids together with 
the F3 and the G2 best fits. 
The $rms$ for the merged light-curve is  $\sim$0.07 (F3) and $\sim$0.08 (G2) mag. 
The coefficients $A_i$ and $\Phi_i$ for the F3 templates and the exponents 
$G_i$, $\Gamma_i$, $\tau_i$ for the G2 templates are listed in Table~\ref{tab_c}

The \textit{IDL} procedure for estimating the mean NIR magnitudes by using the  templates is available on the webpage: \url{http://www.laurainno.com/#!idl/c5wp}.

\subsection{Validation of the new phase zero--point: bump Cepheids}
\begin{figure}[!t]
\begin{center}

\includegraphics[width=0.80\columnwidth] {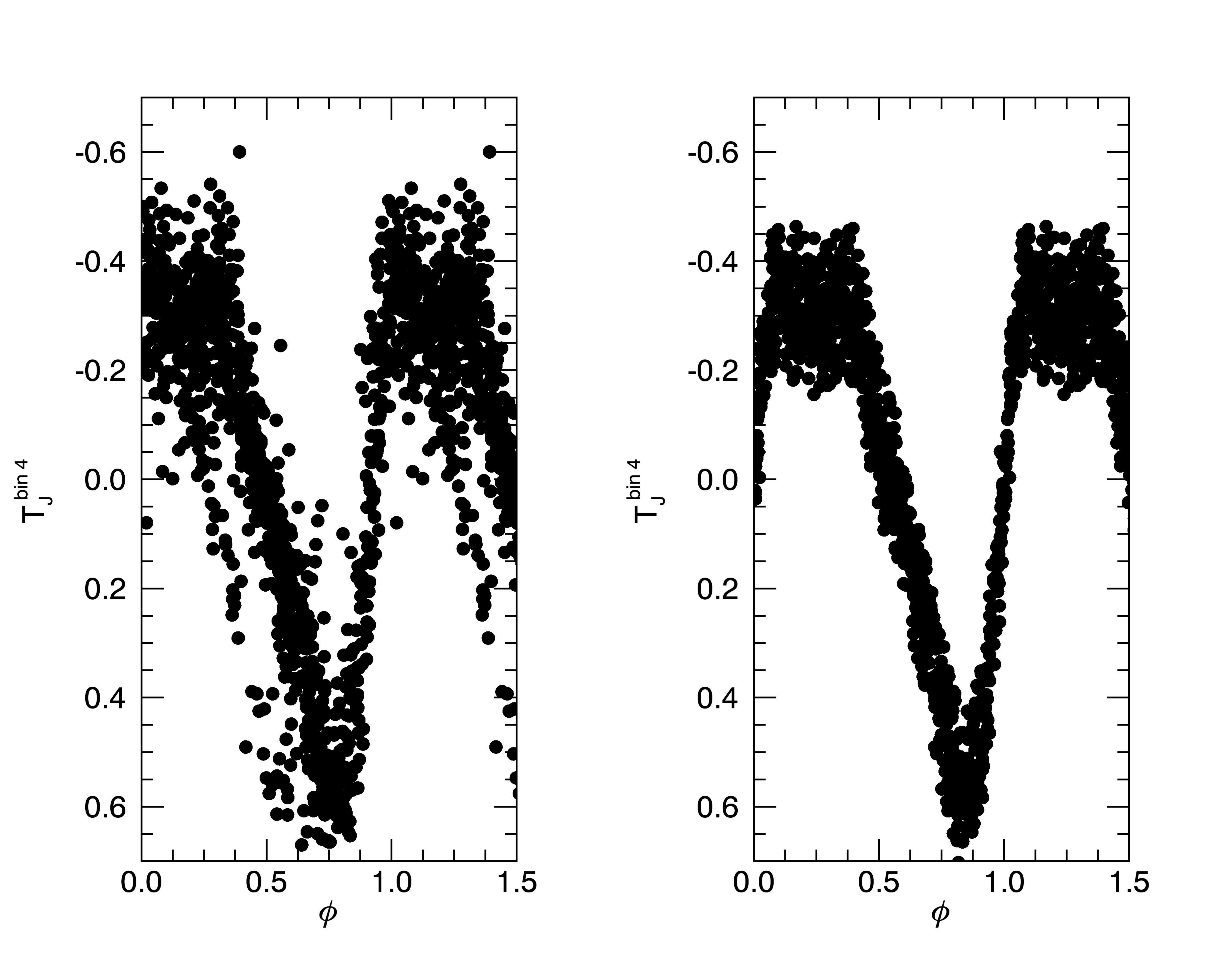}
\caption{Comparison between the merged light-curve  T$^{bin~4}_J$ with phasing anchored 
on the phase of maximum light (Eq. 1; left) and with the new phasing anchored on 
the phase of mean magnitude along the rising branch (Eq. 2; right) for bin~4 
Cepheids.}
\label{f7}
\end{center}

\end{figure}
To improve the mean NIR magnitude of bump Cepheids, we adopted (see Sect. 2.1) 
a new phase zero--point anchored on the phase of the mean magnitude along the 
rising branch. Fig.~\ref{f7} shows the comparison between the merged $J$--band 
light curve for the period bin~4 computed by using as phase zero--point both 
the phase of maximum light (Eq.~1, left panel) and the phase of mean magnitude 
along the rising branch (Eq. 5, right panel). 
A glance at the data plotted in this figure show that the $rms$ significantly 
decreases in the merged light-curve that was computed using the new phase zero--point, 
and the $rms$ decreases by roughly a factor of two (0.06 vs 0.13). 

\begin{figure}[!ht]
\begin{center}
\includegraphics[width=0.80\columnwidth] {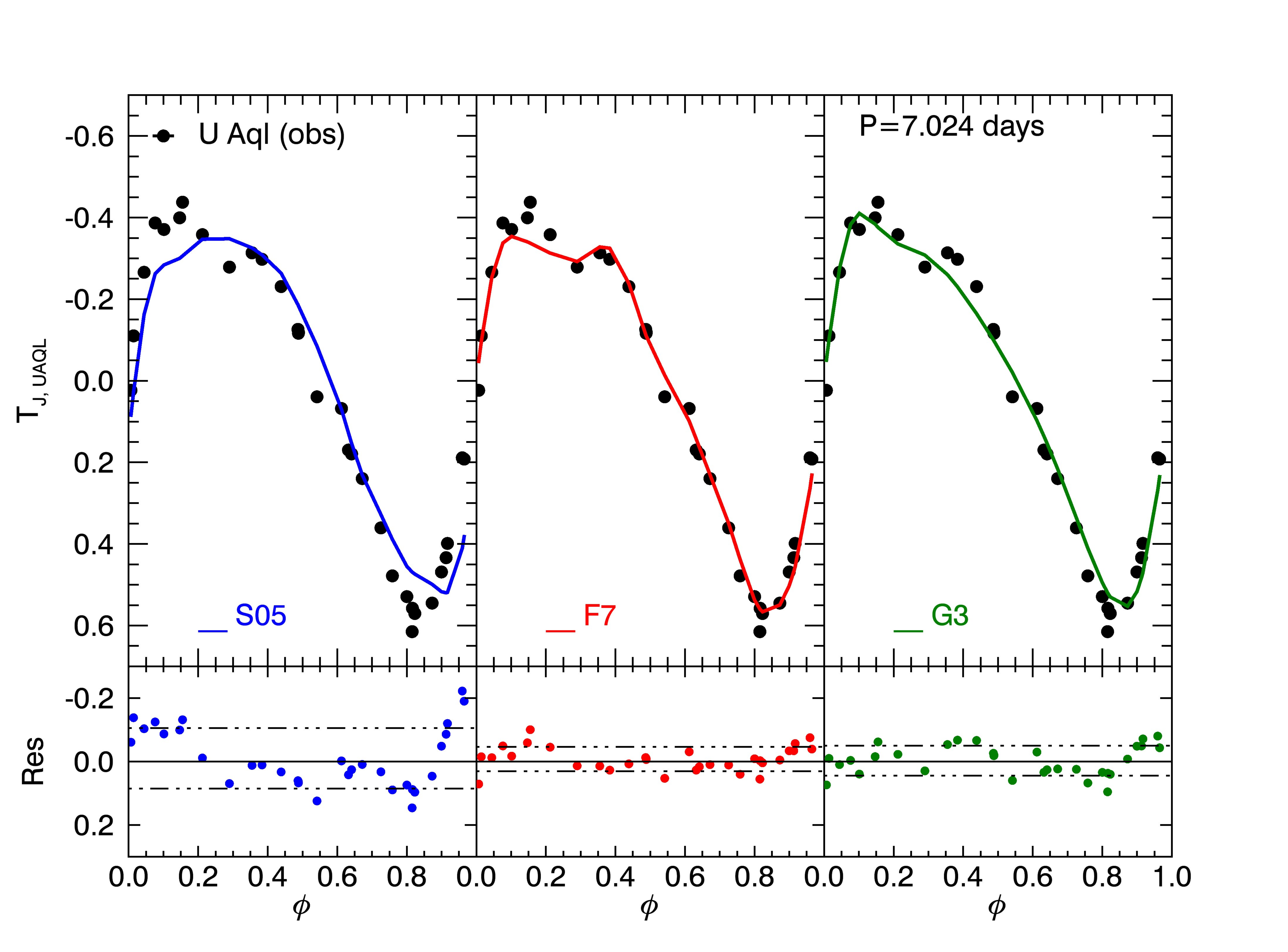}
\caption{From left to right: comparison between the normalized $J$--band light-curve 
(black dots) of U Aql and the  S05 (blue line; left), 
 F7 (red line; center) and  the G3 (green line; right) templates. The typical error 
associated with observations ($\pm$0.01 mag) and rescaled in normalized units is about $\sim$0.001 and is shown in the top left corner of the plot. 
The residuals between the data and the templates are plotted in the bottom panels. 
The dashed lines indicates the $rms$ of the residuals; it decreases from 0.10  (S05) 
to 0.04 (F7, G3).}
\label{f8}
\end{center}
\end{figure}

To further constrain the precision of current templates, we also performed 
a comparison with the light-curve template provided by 
S05\footnote{We applied the S05 templates by using their phase zero--point 
--the epoch of the maximum-- and their NIR--to--optical  amplitude ratios.}. 
The left panel of Fig.~\ref{f8} shows that the $J$--band S05 template
predicts a shape of the normalized light-curve (left panel, blue line) 
for Cepheids with period approaching the center of HP (U Aql, P=7.024) 
that differs from the  observed shape. The difference is quite clear not only 
close to the phase of the maximum ($\phi \sim$0.05), but also close to 
the phase of the bump ($\phi \sim$0.15), and in particular, along the 
decreasing branch.  The middle and right panels show the 
comparison between our templates (F7, middle; G3, right) and the
observed data. 
The residuals for the S05 template plotted in the bottom left panel of the same figure 
display a phase delay between the data and the light-curve template 
along the rising and the decreasing branch. On the other hand, 
the residuals plotted in the middle and right panels
show that  the F7 and G3 templates provide a good approximation 
of the observed light-curve. 
The residuals have an $rms$ (dashed lines) of 0.04 mag, that is a factor of two smaller than 
the $rms$ of the S05 template (0.10).

\section{NIR--to-optical amplitude ratios}
\begin{figure*}[!t]
\begin{center}
\includegraphics[width=0.80\textwidth, height=0.4\textheight] {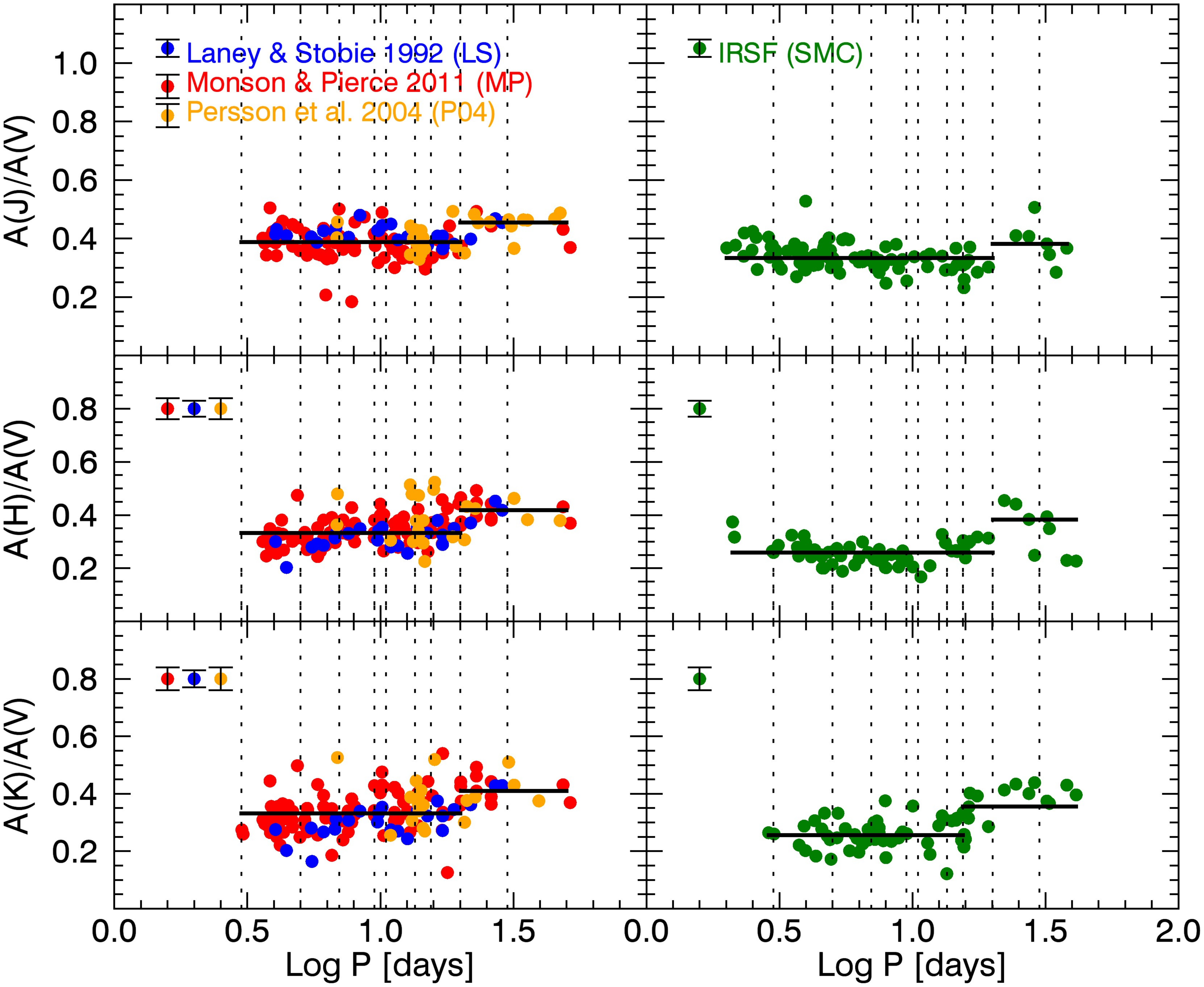}
\caption{Top left: amplitude ratios -- $A_J/A_V$ -- as a function of the period 
for LMC and MW calibrating Cepheids. The dashed vertical lines display the individual 
bins in period. The color coding is the same as in Fig.~1: orange dots for the P04 
sample, red dots for the MP sample, and blue dots for the LS sample. The black solid 
lines are the mean amplitude ratios estimated on the selected period ranges: 
$P \le$ 20 days and $P >$20 days. The error bars in the top left corner show 
the typical photometric error. 
Top right:  Same as the left, but for SMC calibrating Cepheids (green dots). 
The comparison between the value of the mean ratio in the left and in right panels 
shows that amplitude ratios of SMC Cepheids are systematically lower than those of LMC+MW Cepheids. 
Middle: Same as the top, but for the $A_H/A_V$ amplitude ratios. 
Bottom: Same as the top, but for the $A_{Ks}/A_V$ amplitude ratios. 
Note that the period ranges for this band are $P \le$ 15.5 days and $P>$15.5 days.}
\label{f9}
\end{center}
\end{figure*}
The light-curve template allows us to estimate the NIR mean 
magnitudes from single-epoch observations if the amplitude 
in that specific band is already known. Indeed, Eq.~5 gives 

\begin{equation}
<J>_l= J_{i,l}-A_{J,l} \times T_J
\end{equation}
and similar equations for the other NIR bands (see also Eq.~4 in S05). 
To estimate the NIR mean magnitudes, the luminosity amplitudes can be 
estimated by using the luminosity amplitudes in the 
optical bands. We derived new amplitude ratios between optical and NIR 
bands by using our calibrating Cepheids. The results are shown 
in Fig.~\ref{f9} from top to bottom: $A_J$/$A_V$, $A_H$/$A_V$, and 
$A_{Ks}$/$A_V$.  We estimated the mean value (black solid line) 
over two different period ranges: P $\le$ 20 days and P$>$20 days for 
 the MW+LMC (left panels) and the SMC calibrating Cepheids 
(right panels). 

The data plotted in this figure disclose several interesting features 
that need to be addressed in more detail, because these ratios 
are prone to systematic uncertainties. Classical Cepheids are young objects 
and a significant fraction of them are still members of binary systems
\citep{szabados12b}. Their companions are mainly young low--mass stars, 
which meant that they mainly affect the $V$--band amplitude. Moreover, recent accurate 
optical and NIR interferometric \citep{kervella06}, mid-infrared 
\citep{marengo10,barmby11}, and radio \citep{matthews12} measurements indicate 
the presence of a circumstellar envelope around several Galactic Cepheids. 
This evidence implies that the NIR amplitudes might also be affected
by systematic uncertainties and it accounts for a significant 
fraction of the dispersion around the mean values because the photometric 
errors are significantly smaller (see the typical error bars in the 
top left corners). 

Moreover, current theoretical \citep{bono00} and preliminary empirical 
evidence  \citep{paczy00,szabados12a}  indicates that the luminosity amplitudes  
depend on the metal content. 
The $V$--band amplitudes of SMC Cepheids in the short--period range 
(P$\le$ 11 days) are, at fixed period, larger than those of Galactic and LMC 
Cepheids. The difference is caused by the HP dependence on the 
metallicity (see Sect.~2.1). The trend is opposite in the long--period range. 
The difference in the optical amplitude is also clear in the Fourier 
amplitude of Magellanic and Galactic Cepheids 
\citep[see Fig.~5 and Fig.~2 in][Fig.~2]{sos2008,sos2010,matsunaga13}. 

This evidence indicates that solid empirical constraints on the dependence 
of the luminosity amplitudes on metallicity requires accurate information on 
individual metal abundances. However, \citet{genovali13} did not find, within the errors, 
any significant dependence on iron abundance, by adopting 
a homogeneous sample of 350 Galactic and 77 Magellanic Cepheids with 
precise and homogeneous iron abundances. However, this finding is far 
from being definitive, because the number of SMC Cepheids for which accurate iron abundances are 
available is quite limited (19).

The data plotted in the top panel of Fig.~\ref{f9} indicate that the NIR--to--optical  
amplitude ratios of SMC Cepheids (right panel) are smaller over the entire 
period range than the amplitude ratios of MW plus LMC Cepheids 
(left panel). This is the reason why we decided to adopt independent values 
for the NIR--to--optical  amplitude ratios of SMC and MW plus LMC Cepheids.  
The high dispersion in the amplitude ratios 
is mainly due to the $V$--band amplitude distribution in the Bailey 
diagram (amplitude vs period), while the NIR amplitudes show a similar 
distribution, but tighter. 
Current theoretical and empirical evidence indicates that the amplitude 
distribution in the Bailey diagram for Galactic 
Cepheids has the typical V shape  \citep{vangenderen74,bono00,szabados12a,genovali14}, with
the largest luminosity amplitudes 
attained in the short-- ($\log P\le$ 1.0 days) and in the 
long--period ($\log P \ge$ 1.2 days) ranges, while the minimum, at fixed chemical composition, is reached at the center of the HP. 
This peculiar behavior does not allow a straightforward prediction of the NIR amplitude on the basis of the period. 
On the other hand, the NIR--to--optical  amplitude ratios are almost constant for a broad range of periods, as shown in Fig.~\ref{f9}. In particular, we find that the  mean $A(J)/A(V)$, and $A(H)/A(V)$ amplitude ratios are smaller for periods shorter than P$\le$20 days and larger for periods longer than  P$>$20 days. For the SMC mean $A(K)/A(V)$ amplitude ratios we chose a different cut in period: P=15.5 days.
The estimated NIR--to--V amplitude ratios for MW+LMC and SMC calibrating 
Cepheids are listed in Table~\ref{tab_amp}. 

We also note that the typical dispersion for the NIR--to--optical  amplitude ratios 
of the MP sample is almost twice as high as that of the LS sample
(MP: $\sigma$=0.05; LS: $\sigma$= 0.03; $J$-band). 
The main reason for this is the photometric quality of $V$-band light-curves in the MP sample. As already mentioned in Sect.~2, the optical photometry for these Cepheids 
was collected from the literature, with the source data spanning a long time
and coming from different instruments.
Indeed, the typical $rms$ for the MP calibrating $V$-band light-curve is $\sim$0.05 mag,
ten times higher than the typical $rms$ for the LS $V$-band light-curves ($\sim$0.005 mag, see also Sect. 2).
We also performed a test by adopting different mean amplitude ratios for 
each period bin and we did not find any significant improvement in the final results. 

Similar considerations apply to the use of linear regression to fit the
amplitude ratios as a function of the period. A linear relation 
on the entire period range  provides a good approximation 
for the short-period range, where the ratios are almost 
constant, but it underestimates the values in the long-period range. 
On the other hand, by adopting two different relations in the two period ranges, 
the number of parameters will double without significantly improving the template accuracy compared with the two horizontal lines.

\begin{figure}[!h]
\begin{center}
\includegraphics[width=0.90\columnwidth, height=0.8\columnwidth] {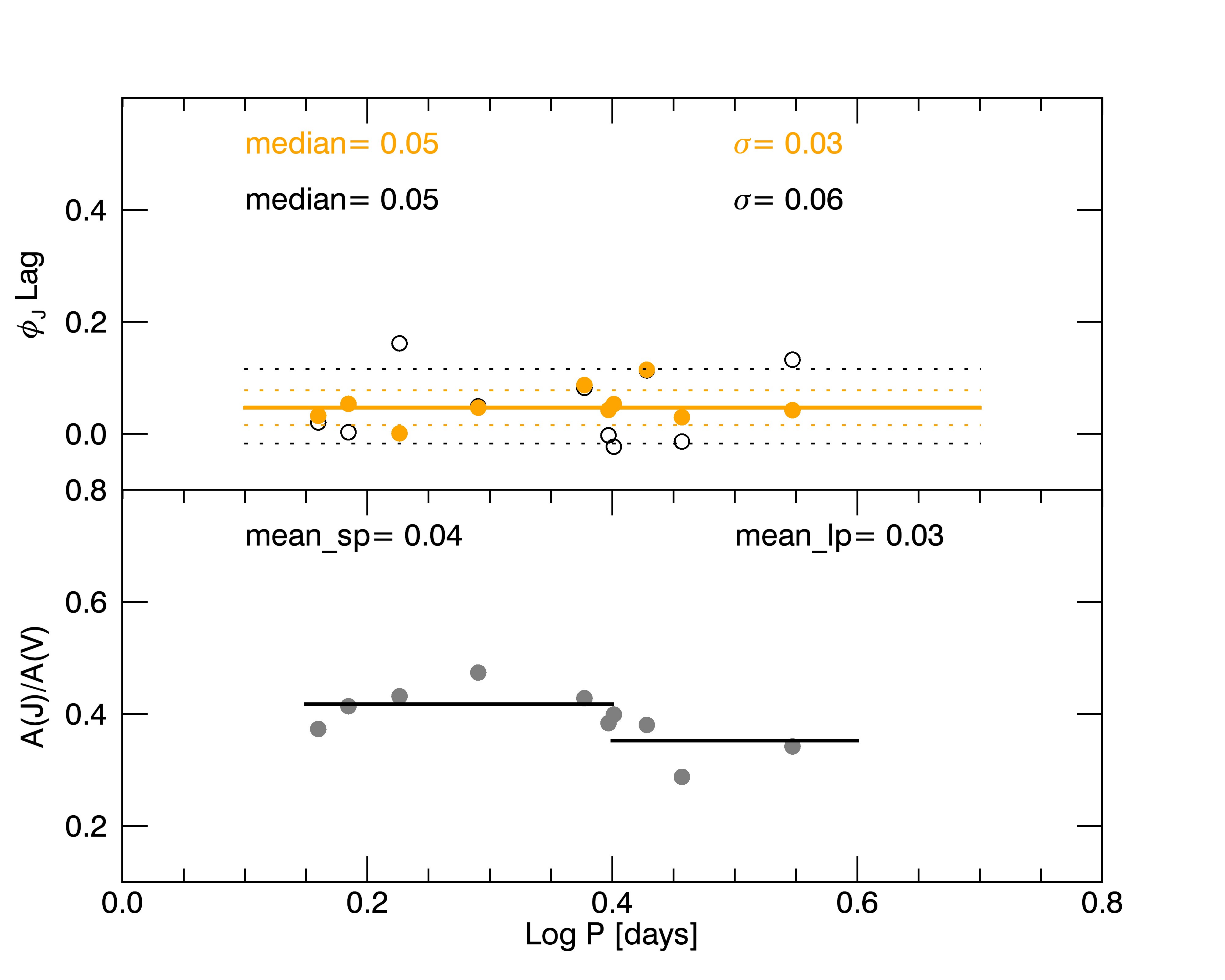}
\caption{Top: Phase lags in $J$ band for FO Cepheids based on the epoch of the 
maximum (black circles) and on the epoch of the mean magnitude along the rising branch 
(orange dots). The median (solid lines) and $rms$ (dashed lines) of the ten 
calibrating Cepheids are overplotted. The black labels refer to the epoch of maximum, 
orange labels to the epoch of the mean magnitude. 
Bottom: the amplitude ratio $A(J)/A(V)$ for FO Cepheids. The mean values for the two 
adopted bins in period (P$\le$2.8, and P$>$2.8 days) are also labeled 
(black solid lines).}
\label{f10}
\end{center}
\end{figure}

The pulsational amplitudes and Fourier parameters of FO Cepheids also show 
a sudden jump for periods close to P=3 days \citep{kienzle99}. This behavior is 
associated with the possible presence of a short-period bump along the light-curves 
of FO Cepheids with periods between 2 and 3.8 days \citep{bono00}.  
Thus, we adopted two different $A(J)/A(V)$ mean values 
for the two different regimes of the amplitude before and after 
the appearance of this short-period bump. 
Fig.~\ref{f10} shows the phase lags (top panel) and the 
amplitude ratios (bottom panel) computed for the FO calibrating Cepheids.
The mean values (solid lines) are $A_J$/$A_V$= 0.40  
(P$\le$2.8 days) and $A_J$/$A_V$= 0.30 (P$>$2.8 days). 


\section{Validation of the light-curve templates}
\begin{figure}[!h]
\begin{center}
\includegraphics[width=0.99\columnwidth]{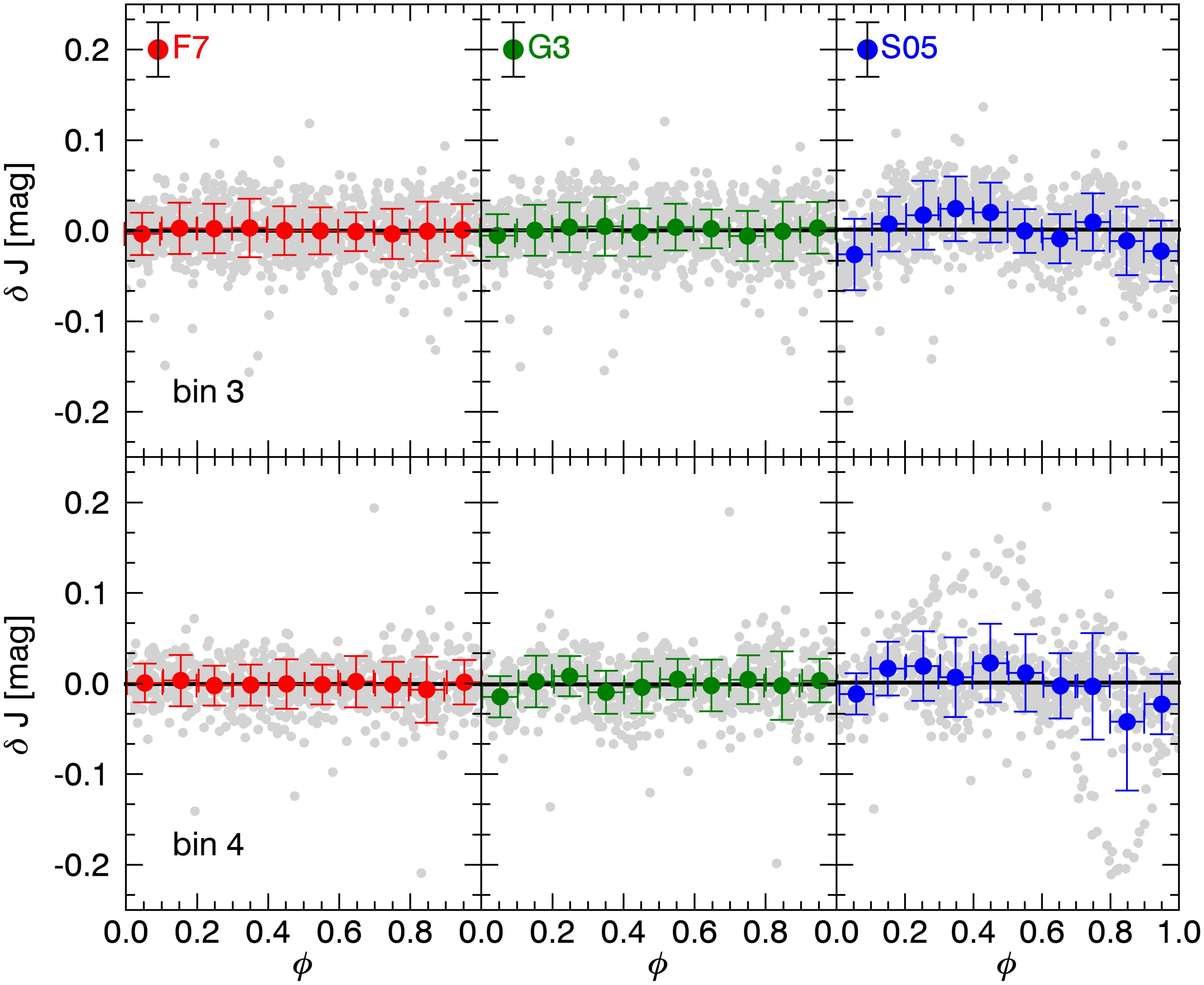}
\caption{Top: $\delta J$ for period bin~3 (5--7 days) by adopting the 
F7 (red dots, left panel), the G3 (green dots, central panel), and the S05 
templates (blue dots, right panel). The gray dots on the background are the 
difference between the mean magnitude estimated by applying the templates 
and the true mean magnitude estimate by the individual fits. By binning the 
phase in ten different bins, we estimated the mean of the residuals 
(dots) and the standard deviation (error bar). The solid black lines show 
the mean values, which are $<$10$^{-3}$ mag for the three templates. 
Bottom: Same as top, but for period bin~4 (7--9.5 days).}
\label{f11}
\end{center}
\end{figure}

To further evaluate the accuracy of current templates, we performed a 
new test by using the complete light-curves of the calibrating Cepheids. 
For each period bin, we have several calibrating light-curves (see Fig.~\ref{f1}) 
for which all the parameters  --mean magnitudes, NIR luminosity amplitudes, 
period, phase zero--points-- have already been estimated. Therefore, we 
randomly selected a phase point from the calibrating light curve
to simulate a single-epoch observation and applied the new templates 
to estimate the mean magnitude, which we then compared with the true one.  
We define $\delta J$ as the difference in $J$ band between the true mean magnitude 
--estimated as the mean along the light-curve-- and the mean magnitude 
computed by using the new $J$--band templates. 
A similar approach was also adopted for the $H$ and $K_{\rm{S}}$--bands. 
Fig.~\ref{f11} shows the $\delta J$ for two period bins: bin~3 (top) 
and bin~4 (bottom) by adopting the F7 (red dots, left panels), the G3 
(green dots, middle panels) and the S05 (blue dots, right panels) 
light-curve templates. The  $\delta J$ based on 
both F7 and G3 templates give a vanishing mean ($\le 10^{-3}$ mag) and 
a small standard deviation $\sim 0.03$ mag. The S05 templates also give 
a mean close to zero ($\sim 10^{-3}$ mag) and a slightly larger standard 
deviation ($\sim 0.04$ mag). The data plotted in Fig.~\ref{f11} show that the residuals
of the S05 template are not symmetric, therefore we estimated the interquartile range 
and found that the difference becomes about 40\% ($\sim$0.06 vs $\sim$0.04 mag).
We also divided the data into ten phase bins
and estimated the mean and standard deviation for each bin.  
The values are overplotted in Fig.~\ref{f11} (red dots, F7; 
green dots, G3; blue dots, S05). The horizontal error bars display the 
range in phase covered by individual bins, while the vertical error bars 
display their standard deviations. 

\begin{figure}[!ht]
\begin{center}
\includegraphics[width=0.90\columnwidth, height=0.8\columnwidth] {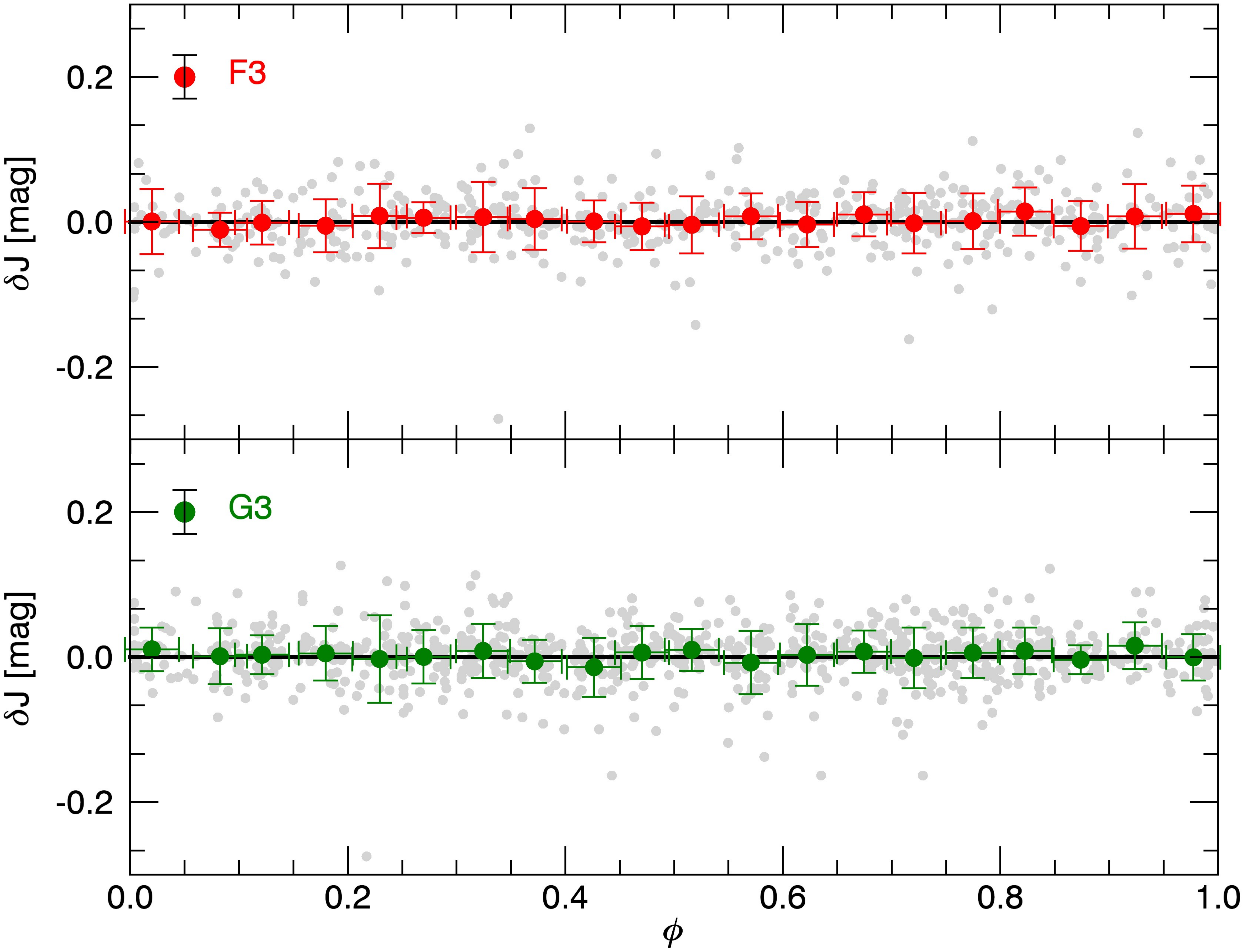}
\caption{Same as Fig.~11, but for FO calibrating Cepheids. 
The $\delta J$ are based on 
the F7 (top) and the G2 (bottom) templates. The red (F3) and green (G2) dots indicate 
the mean for each of the 20 bins in phase and the standard deviation (error bars). 
The mean of the residuals is $\le 0.01$, with a mean $rms$ of 0.04 mag.}
\label{f12}
\end{center}
\end{figure}

The data plotted in this figure indicate 
that the residuals of the  F7 and G3 are independent of the phase, while the 
residuals of the S05 template show a clear phase dependence for 
$\phi\sim$0.5 and $\phi\sim$0.8. In particular, the mean $J$ magnitudes 
based on the S05 template are 2$\sigma$ fainter close to 
$\phi\sim$0.5 and brighter close to $\phi\sim$0.8. Most of this discrepancy 
is due to bump Cepheids for which the phase zero--point anchored on the 
epoch of maximum brightness introduces systematic phase shifts in using the template. 
Similar trends were also found when estimating the $\delta H$ and 
$\delta K_{\rm{S}}$ by applying the F7, G3, and S05 templates. 
We also performed the same test for FO Cepheids, and the results are plotted 
in Fig.~\ref{f12}. Once again, the residuals for the F7 and the G3 templates 
attain vanishing values ($\le 10^{-3}$) and the standard deviations are 
smaller than 0.04 mag.  
Note that above standard deviations account for the entire error budget, 
because they include the photometric error (measurements, absolute calibration) 
and the standard deviations of the analytical fits. 

\subsection{Test based on single-epoch measurements}

The validation of the light-curve templates performed in the last section 
has a limited use, bceause it relies on Cepheids with a good coverage 
of the pulsation cycle, and in turn on accurate luminosity amplitudes. 
Therefore, we performed an independent validation by using the amplitude 
ratios discussed in Sect.~5 and by randomly extracting phase points from 
the $J$, $H$, and $K_{\rm{S}}$--band light-curves of the calibrating 
SMC Cepheids. This is an acid test, because we are mimicking the typical
use of the light-curve template.  

\begin{figure}[!t]
\begin{center}
\includegraphics[width=0.99\columnwidth, height=0.8\columnwidth]{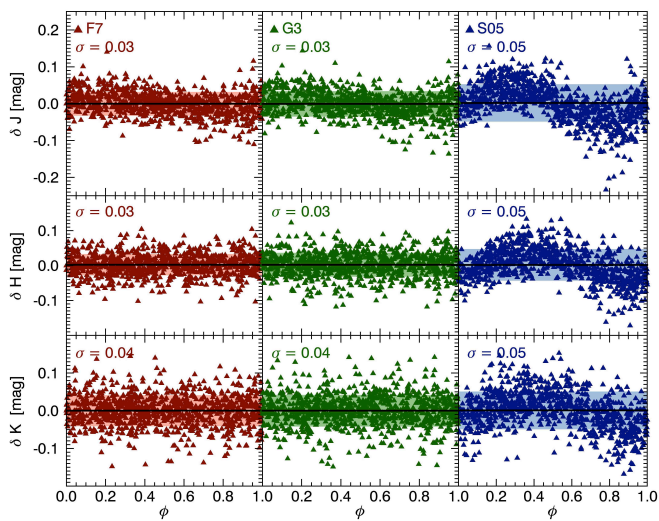}
\caption{Random-phase extraction test for all the period bins in the different NIR bands 
(from top to bottom: $J$, $H$ and $K_{\rm{S}}$) in the SMC sample. The red triangles 
show the difference between the mean magnitude predicted by applying the F7 template 
to the single-epoch NIR observation (left panel). In the middle panel the green triangles 
were evaluated by applying the G3 template, while the blue triangles in the left panel 
are evaluated by applying the S05 template. The shadowed areas shows the 2$\sigma$ of 
the results (F7: light red; G3: light green; S05: light blue), while the black lines 
indicate the zero. The residuals of F7 and G3 template show no dependence on the phase, 
and the dispersion is from $\sim$20\% ($K_{\rm{S}}$) to 40\%($J$) lower than the 
residuals of the S05 template.}
\label{f13}
\end{center}

\end{figure}

The difference in mean magnitude between the true mean magnitudes and 
the mean magnitudes estimated using the new NIR templates are plotted in 
Fig.~\ref{f13} as red triangles (F7, left panels), green triangles 
(G3, middle panels), and blue triangles (S05, right panels). The red, 
green and light blue shadowed areas indicate the standard deviation
$\pm \sigma$, for the $\delta J$ (top), the $\delta H$ (middle) and the $\delta K_{\rm{S}}$ 
(bottom) estimated with the three different templates.
The  mean values for the three bands ($\delta J$, $\delta H$ and 
$\delta K_{\rm{S}}$) are lower than a few thousandths in all the cases. 
However, the $\sigma$ for the F7 and the G3 $J$ and $H$--band 
templates are at least 40\% lower than for the S05 template 
(0.03 vs 0.05 mag). The difference for the $K_{\rm{S}}$ band is lower 
and of about $\sim$20\%  (0.04 vs 0.05 mag). 
Moreover, the residuals of the S05 template show a phase dependence 
that is not present in the residuals of the F7 and the 
G3 templates\footnote{Note that for this test we only adopted calibrating 
Cepheids with $\log P \geqslant$0.5, because the S05 template does not include 
shorter period Cepheids.}. 

The evidence that the new NIR templates do not significantly reduce the 
scatter in the $K_{\rm{S}}$ band is a consequence of the fact that the pulsation 
amplitude in this band is smaller than in the $J$ and in the $H$ bands. Moreover,  
the photometric errors on individual measurements become larger.   

\begin{figure}[!t]
\begin{center}
\includegraphics[width=0.99\columnwidth, height=0.8\columnwidth]{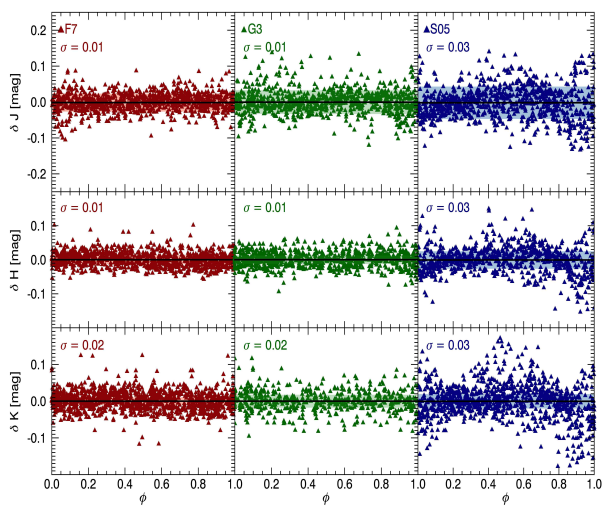}
\caption{Same as Fig.~\ref{f13}, but for Galactic Cepheids (LS, MP). The red triangles 
show the difference between the mean magnitude predicted by applying the F7 template 
(left panel), the G3 template (central panel, green triangles), and the S05 template 
(right panel, blue triangles). The shadowed areas shows the 2$\sigma$ of the results 
(F7: light red; G3: light green; S05: light blue), while the black lines indicate the 
zero for each data set.}
\label{f14}
\end{center}
\end{figure}

We performed the same test for the Galactic calibrating Cepheids, and the 
results are given in Fig.~\ref{f14}. 
The difference we found for Galactic calibrating Cepheids 
from using the new NIR templates is between two ($K_{\rm{S}}$) to three ($J,H$) 
times smaller than for the S05 templates. The mean in the three different 
bands approaches zero ($\sim10^{-3}$ mag), but the residuals of the 
S05 template show a clear phase dependence. 
Moreover, data plotted in the right panels show a systematic 
overestimate of the mean magnitude for phases close to the rising branch
($\phi\sim$1). The main reason for this discrepancy is once again the 
adopted phase zero--point. The use of the maximum brightness to anchor the 
phase causes loose constraints along the rising branch, i.e. close to phases 
in which the luminosity rapidly increases. The F7 and the G3 templates 
do not show evidence of similar systematic effects. However, the former 
exhibits a mild phase dependence in the $K_{\rm{S}}$ bands --and probably in the 
$H$ bands-- close to $\phi$=0.8. 

The difference between the residuals of MW+LMC and SMC calibrating 
Cepheids is a consequence of the fact the former sample is characterized 
by a better photometric precision over the entire period range.  
The mild phase dependence is mainly due to the smaller photometric errors, 
and in turn, to the smaller standard deviations (see labeled values). 
The key points to explain the above trends are 
a) the definition of the phase zero--point: the use of the mean magnitude 
along the rising branch (Sect.2.1) instead of the decreasing branch reduces the 
precision of the template for periods longer than $13.5$ days; 
b) the NIR--to--optical  amplitude ratios adopted for the $H$ and $K_{\rm{S}}$: 
the $V$--band amplitude of the Galactic calibrating Cepheids shows, at fixed period, 
a high dispersion, that propagates into the NIR--to--optical  amplitude ratios. 

\begin{figure}[!h]
\begin{center}
\includegraphics[width=0.90\columnwidth, height=0.8\columnwidth] {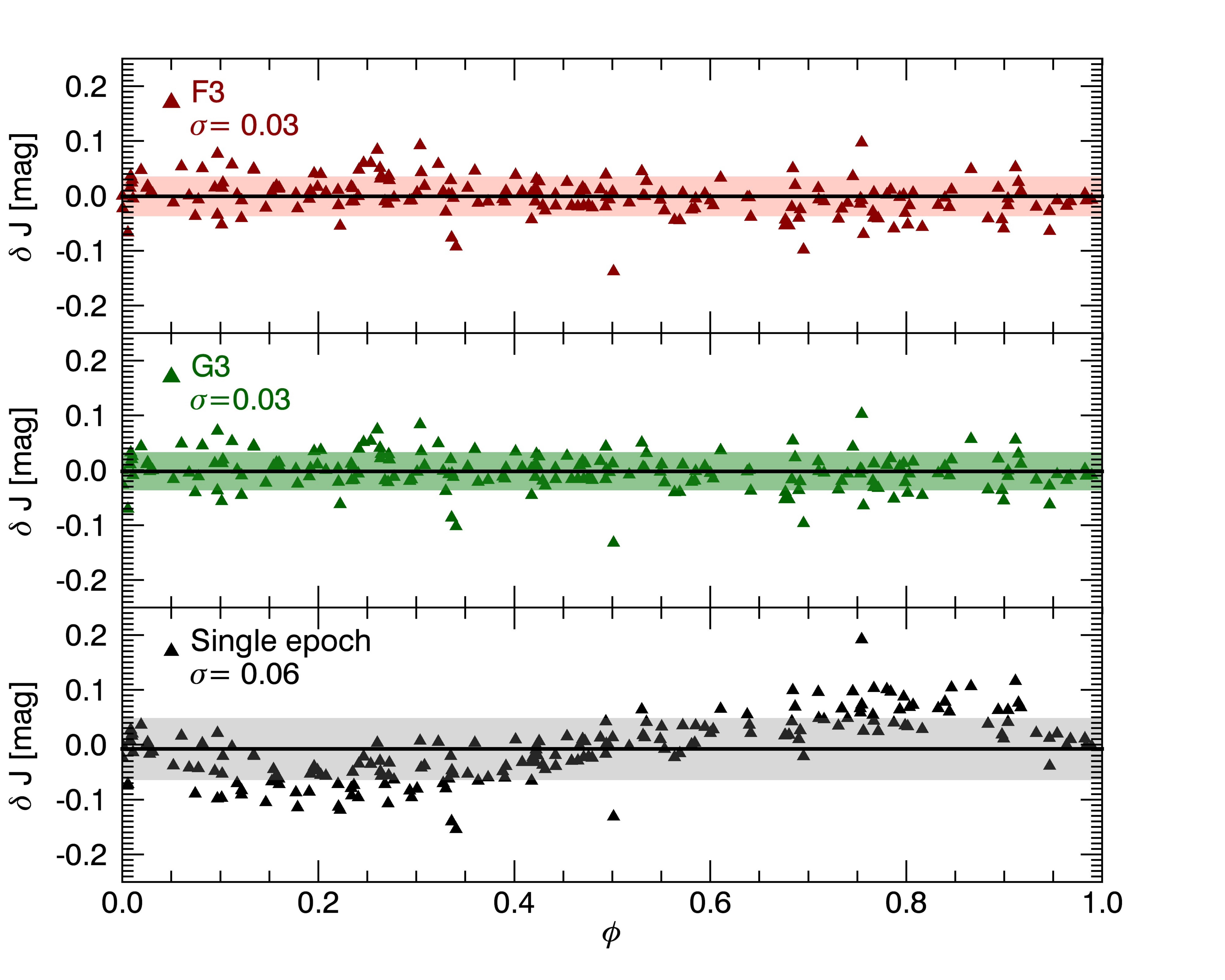}
\caption{Random-phase extraction for FO calibrating Cepheids in the $J$--band. 
The red triangles show the differences between the mean magnitude based on the 
F3 template and on single--epoch  NIR observations (top panel). The middle panel 
shows the residuals (green triangles) based on the G2 template, while the 
bottom panel shows the residuals (black triangles) of the single--epoch measurements. 
The shadowed areas shows the 2$\sigma$ of the results (F7: light red; G2: light green; 
single-epoch: light gray), while the black lines display the mean. }
\label{f15}
\end{center}
\end{figure}

The same test was also applied to the FO calibrating Cepheids. 
The residuals plotted in Fig.~\ref{f15} clearly show that the standard 
deviation of the $J$--band template decreases by $\sim$50\% when 
compared with single-epoch measurements extracted along the light-curves.
Note that the typical pulsation amplitude in the $J$ band 
for FO pulsators is $\sim$0.15 mag. This means that the use of single-epoch 
measurement as a mean magnitude introduces an error of about
$\sim$0.07 mag.  Thus, the new FO NIR templates allow us to reduce the error 
budget for FO Cepheids by almost a factor of two. Moreover, the new templates 
do not show evidence of a phase dependence.  

To fully exploit the impact of the new NIR templates on FU Cepheids, 
we also performed a test with the mean Wesenheit magnitudes. 
The NIR Wesenheit magnitudes are closely related to apparent magnitudes, 
but they are minimally affected by uncertainties on reddening and are 
defined as

$$W(JK_{\rm{S}})=K_{\rm{S}}-0.69 \times (J- K_{\rm{S}}),$$
$$W(HK_{\rm{S}})=K_{\rm{S}}-1.92 \times (H- K_{\rm{S}}),$$
$$W(JH)=H-1.63 \times (J- H),$$
where the coefficients of the color terms are based on the reddening law provided 
by \citet{cardelli89}, and for the SMC selective absorption coefficient we adopted 
$R_V$=3.23. This is an acid test concerning the NIR templates, since the color 
coefficients of the Wesenheit relations attain values higher than one for bands 
with limited difference in central wavelength. This means that uncertainties 
affecting the mean colors are magnified in using Wesenheit relations.    

\begin{figure*}[!ht]
\begin{center}
\includegraphics[width=0.8\textwidth, height=0.4\textheight]{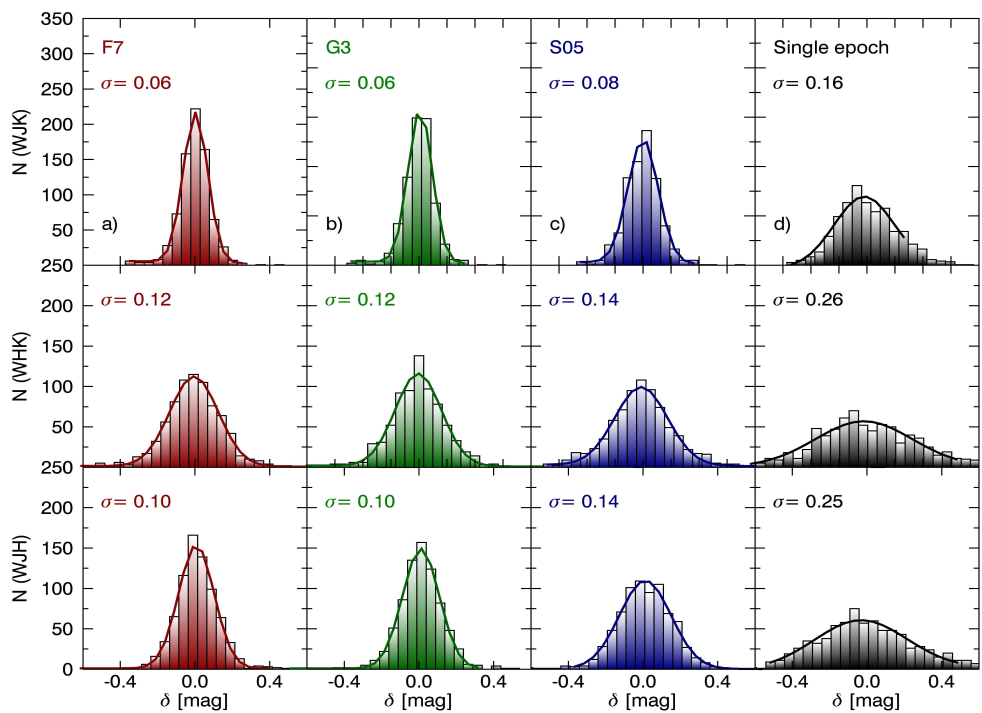}
\caption{Top: From left to right: histogram of the difference between 
the true mean Wesenheit $W(JK_{\rm{S}})$ and the mean Wesenheit magnitude 
estimated by using the F7, G3, S05 templates, and the single-epoch data. 
The single-epoch data were randomly extracted from the calibrating SMC light-curve  
at different phases for the $J$, $H$, and the $K_{\rm{S}}$ bands. 
 The Gaussian fits from which the labeled standard deviation has been estimated are also overplotted on the histograms.
Middle: Same as top, but for the W($HK_{\rm{S}}$) magnitude. 
Bottom: Same as top, but for the W($JH$) magnitude.}
\label{f16}
\end{center}
\end{figure*}

To simulate the effect of non-simultaneous NIR observations, we randomly 
extracted for the entire set of SMC calibrating light-curves three different 
($J$,$H$,$K_{\rm{S}}$) measurements. The NIR templates were applied to each of 
them and we obtained three independent estimates of the mean NIR magnitudes. 
Then we computed the three mean Wesenheit magnitudes 
--$W(JH)$,$W(HK_{\rm{S}})$, $W(JK_{\rm{S}})$-- by using these relations and 
estimated the difference in magnitude with the true mean Wesenheit magnitudes. To properly 
sample the luminosity variations along the entire pulsation cycle, the 
procedure was repeated ten times per light-curve. 
This test was performed with the new (F7, G3) and the S05 templates. 
The residuals are plotted in Fig.~\ref{f16}, $W(JK_{\rm{S}})$ (top), 
$W(HK_{\rm{S}})$ (middle) and $W(JH)$ (bottom). 
 A Gaussian fit to the histograms performed to evaluate the mean and the standard deviation is also overplotted. We found that the means 
once again vanished. 
The $\sigma$ of the mean Wesenheit magnitudes based on the F7 [a) panels] and 
on the G3 [b) panels] templates are between 15--30\% lower than the $\sigma$ 
of the residuals based on the S05 [c) panels] template (see labeled values). 

Finally, we also compared the difference between the mean Wesenheit magnitudes 
based on the new NIR templates with single-epoch measurements randomly extracted 
from the light-curve of the SMC calibrating Cepheids. The gray shaded areas 
plotted in the d) panels of Fig.~\ref{f16} show that the standard deviations 
of the new NIR templates are a factor of two smaller than those of the 
single-epoch measurements. 

\subsection{Error budget of the analytical fits}

These results clearly show that the application of NIR 
light curve templates increases the accuracy on the mean magnitude compared 
with single-epoch measurements. However, the templates are affected by several 
uncertainties that contribute to the total error budget. 
The test discussed in Sect. 6.1 and shown in Fig.~\ref{f13} was also applied 
to constrain the impact of the individual uncertainties on the total dispersion 
of the $\delta J$, $\delta H$, and $\delta K_{\rm{S}}$ residuals.

{\sl i) Photometric error} -- The photometric error is the main source of error, 
and it affects the precision of the template and its application. However, only 
the former source should be taken into account when estimating the precision 
of current templates. To artificially remove the photometric error on the 
measured NIR magnitudes, we extracted the individual measurements from the 
Fourier fits of the light-curves. The result of this numerical experiment 
shows that 60\% of the total dispersion is due to the photometric error 
on the observed magnitudes. This accounts for 0.02 mag in the total error budget.   

{\sl ii) Use of the template} -- The use of a template plus a single-epoch 
measurement to estimate the mean magnitude accounts for 12\% of the total 
dispersion in Fig.~\ref{f13}. 

{\sl iii) Merging of the light-curves in period bins} --  Our approach assumes 
that all the light-curves inside the same period bin are identical within the errors. 
This assumption is verified inside the confidence level given by 
the $rms$ of the merged light-curves, typically $\sim$0.03 mag. However, small 
differences in shape may occur between the true light-curve of the Cepheid and 
the given template. This is a simple consequence of the merging process of 
the Cepheid light-curves in a limited number of period bins. We tested the impact 
of this approach by using synthetic light-curve based on analytical fits 
(F7, G3). We found that 15\% of the total dispersion is due to the process 
of merging the light-curves in a modest number of period bins.

{\sl iv) $V$-NIR amplitude ratio} --  The prediction of the pulsation amplitudes 
in the $J$, $H$, and $ K_{\rm{S}}$ bands based on the optical amplitude introduces 
an uncertainty on the mean magnitude provided by the templates. However, we can quantify this effect 
by adopting the true amplitudes measured for the calibrating Cepheids. The error 
associated with the use of the  $V$-NIR amplitude ratios given in Sect.~5 accounts for 
10\% of the total budget. 

{\sl v) $V$-NIR phase lags} --  We estimated the epoch of the mean magnitude along the rising 
branch by adopting $V$-NIR phase lags (Sect.~2.1). The comparison between the $\delta J$, 
$\delta H$, and $\delta K_{\rm{S}}$ residuals estimated with the adopted and the measured 
epoch of the mean magnitudes indicates that this assumption accounts for $\sim$3\% of the 
error budget.

The error associated with the NIR mean magnitudes estimated by applying the new NIR templates is $\sim$0.015 mag for the $J$ band, 0.017 mag for the $H$ band and 0.019 mag for the $K_{\rm{S}}$ band. 
These errors have to be added in quadrature to the photometric error on the single-epoch 
measurements to which the template is applied. The use of two or more measurements for the 
same Cepheid and the weighted average of the independent mean magnitudes implies a better 
precision on the final mean magnitude.

\section{Summary and final remarks}

We developed new NIR $J$, $H$, and $K_{\rm{S}}$ light-curve templates for FU and FO Cepheids. 
The new templates compared with those already available in the literature have several 
advantages: 

{\sl i)} {\bf Period binning} -- We divided the entire period range (1--100 days) into 
ten different period bins. The binning was performed to
a) reduce the $rms$ of the merged light-curves, 
b) properly trace the change in the shape of the light-curve across the HP, and to 
c) minimize the discrepancy in the amplitude ratio and in the phase difference 
($R_{l,m}$, $\phi_{l,m}$; Fourier parameters) between templates and individual 
light curves. The adopted binning in period allowed us to limit this
difference to less than 15\% of the total error associated with the estimate of 
the NIR mean magnitudes.

{\sl ii)} {\bf Phase zero--point} -- The phase zero--point of the mean magnitude was 
fixed along the rising branch of the light-curve. The main advantage of this new 
definition is that it allow us to estimate the phase lag between optical and 
NIR light-curves. Moreover, the identification of the new phase zero--point 
is straightforward even for bump cepheids and thus overcomes possible systematic 
errors introduced by the secondary bumps along the light curves. 

{\sl iii)} {\bf NIR--to--optical amplitude ratios} -- To apply the new templates, 
we need to know the luminosity amplitude in an optical band 
($V$, $B$) in advance together with the ratio between optical and NIR bands. The optical amplitudes come from the OGLE data set. We provided a new estimate of the 
$V$--NIR amplitude ratios for the calibrating Cepheids and found that
a) the ratios for SMC Cepheids are, at fixed period, systematically lower 
than the ratios of Galactic and LMC Cepheids; 
b) the difference between SMC and MW$+$LMC decreases for periods longer 
than the center of the HP. The optical amplitudes of Galactic Cepheids are 
smaller than the amplitudes of SMC Cepheids for periods shorter than the center 
of the HP. 
Therefore, we adopted two different ratios for the short-- and the 
long--period regime. 

{\sl iv)} {\bf Analytical Functions} -- Together with the popular seventh-order 
Fourier series fitting, we also provided a template based on multi-Gaussian 
periodic functions. The main advantage in using this new template is that 
it provides a solid fit of the light-curves by using fewer
parameters than the Fourier fit. Moreover, it is less sensitive to spurious features that can be introduced in the light-curves by photometric errors and to secondary features (bumps, dips) that can appear along the light-curves.

{\sl v)} {\bf FO pulsators} -- We provide for the first time the $J$-band 
template for FO Cepheids, following the same approach as we adopted for 
FU Cepheids. The new template reduces the uncertainty 
on the mean $J$--band magnitude of FO Cepheids by a factor of two.

The application of the new NIR templates when compared with single-epoch NIR 
data provides mean magnitudes that are 80\% more accurate, 
and their typical error is smaller than $0.02$ mag.
Cepheids mean magnitudes can be used to estimate
their distances by adopting Period--Luminosity (PL)
and Period--Wesenheit (PW) relations.
The error associated with these distances includes
both the error on the observed mean magnitude 
and the uncertainty on the absolute magnitude 
estimated by the adopted relation. 
This uncertainty is formally derived by the dispersion of the relation, 
which is produced by three different error sources:
the photometric errors associated with the measured mean magnitudes
from which the adopted PL or the PW relation is derived, 
the line-of-sight depth of the galaxy, and the intrinsic scatter.
This last term is a consequence of the fact that
PL and PW relations do not account for all the physical parameters that contribute
to the stellar luminosity, such as temperature, metallicity, helium-content.
Recent theoretical predictions \citep{bono00,marconi05,fiorentino07}
and empirical results  \citep{bono10,inno13}
indicate that the intrinsic dispersion decreases for NIR PW relations.
In particular, \citet{fiorentino07} predicted a dispersion lower than 0.05 mag for 
PW$(J,K)$ relation.  
This uncertainty of $\sim$ 3\% on individual distances is thus the
precision that intrinsically limits the method we adopted. 
The main advantage of the new templates is that 
they reach the precision limit, 
even with single-epoch NIR observations.
Indeed, for single-epoch measurements with 1\% accuracy or better, 
the error on the mean magnitude is lower than 2\%.
Computing the sum in quadrature of all these error sources,
the dominant term is then the intrinsic scatter of the PW
relation. 

This means that Cepheid distances can be determined
with the highest possible accuracy by using the new templates 
and one single-epoch NIR observation.  

Compared with the S05 templates, F7 and G3 templates have 
the advantage to be minimally affected by problems 
in phase dependences, and they provide new NIR mean magnitudes 
that are more accurate by 30\%($K_{\rm{S}}$) to 50\% ($J$). 
This means that if single-epoch measurements 
are available with photometric precision better than 0.03 mag,
the new templates already reduce the total uncertainty 
on distances by 20\%  with respect to the S05 templates.
For instance, by applying the new templates to the NIR
single-epoch data presented in \citet{inno13} for the SMC Cepheids, 
the total dispersion of the optical--NIR  
PW relations decreases by up to 30\% (i.e. 0.15 mag vs 0.26 mag, PW$(VJ)$)
when compared to single-epoch data, 
and up to 5\%  (i.e. 0.15 mag vs 0.16 mag PW$(VJ)$) when compared with the S05 template. 
Moreover,the total dispersion for PW$(VJ)$ is 0.15 mag, 
indicating that the scatter due to spatial effects is still 
significantly larger then the intrinsic dispersion ($\sim$3 times larger).
This means that Cepheid relative distances 
can be safely used to derive the three-dimensional structure of the SMC,
with an accuracy limited by the total error estimated above that is $\le$5\%, 
which corresponds to the physical limit of the method itself. 

If we instead consider all of the Cepheids as a statistical ensemble representing 
the stellar distribution in the galaxy, the mean distances to the SMC as derived from different PW relations can be determined with a precision of up to ~0.1\% (0.002 magnitudes, PW$(VJ)$), with the precision scaling as the square root of the number of stars in the ensemble itself ($\gtrsim$2,200 Cepheids).

A more detailed discussion on the application of the new templates
to derive new MC Cepheids relative distances will be given 
in a forthcoming paper (Inno et al., in preparation).

Our findings rely on a panoply of Galactic and MC Cepheid light-curves.  
The new templates and reddening-free optical--NIR PW relations will provide 
accurate absolute and relative distances. The latter appear very promising, 
because the intrinsic error is on the order of 1--2 percent. This gives the 
opportunity to derive 3D structure of nearby stellar systems by using 
single-epoch NIR observations.

In spite of the substantial improvement in the intrinsic accuracy of the 
NIR light-curve templates, the observational scenario is far from being 
complete. Current NIR light-curves did not allow us to derive accurate $H$- and $K_{\rm{S}}$-band templates for FO Cepheids, because of the limited photometric 
accuracy in the short-period regime (faintest Cepheids).
Moreover, we found evidence that the NIR--to--optical  
amplitude ratios of SMC Cepheids are lower when compared with MW+LMC 
Cepheids. Current data did not allow us to constrain whether a similar difference is present in addition between MW and LMC Cepheids, because of the limited sample of 
LMC Cepheids, 
The new NIR time series data that are being collected by IRSF for MC Cepheids 
appear a very good viaticum to address these open problems.  

\begin{acknowledgements} 
This work was partially supported by PRIN--INAF 2011 ``Tracing the
formation and evolution of the Galactic halo with VST" (P.I.: M. Marconi)
and by PRIN--MIUR (2010LY5N2T) ``Chemical and dynamical evolution of
the Milky Way and Local Group galaxies" (P.I.: F. Matteucci).
One of us (G.B.) thanks The Carnegie Observatories visitor program 
for support as science visitor.
N.M. acknowledges the support by Grants-in-Aid for
Scientific Research (Nos. 23684005 and 26287028)
from the Japan Society for the Promotion of Science (JSPS).
Support from the Polish National Science Center grant 
MAESTRO 2012/06/A/ST9/00269 is also acknowledged.
WG gratefully acknowledges support for this work from the BASAL
 Centro de Astrof\'{i}sica y Tecnolog\'{i}as Afines (CATA) PFB-06/2007, and
from the Chilean Ministry of Economy, Development and Tourism's
Millenium Science Iniciative through grant IC120009 awarded to the
Millenium Institute of Astrophysics (MAS).\\
We also acknowledge G. Fiorentino for many useful discussions 
concerning the theoretical predictions on NIR Period Wesenheit
relations.
It is also a pleasure to thank an anonymous referee for his/her
supportive attitude and insightful suggestions that helped us to improve
the readability of the paper.
\end{acknowledgements}




%
\clearpage

\onecolumn
\scriptsize
\begin{landscape}
\begin{longtable}{lclrrrrrrrrcl}
\caption{\label{tab_cat_fu} Pulsation parameters for FU calibrating Cepheids.}\\
\hline\hline 
ID$^{a}$& ID$^{b}$& $\log P$ & $<V >$ & $<J>$ & $<H>$ & $<K_S>$ & $A_V $ & $A_J$ &  $A_H$ & $A_K$ &JD$_{mean}^V$ & Sample\\  
  &  & [days] & [mag] & [mag] & [mag] & [mag] &[mag] & [mag] & [mag] &[mag] &[HJD] & \\
\hline 
  \endfirsthead \caption{continued.}\\ \hline\hline
  ID$^{a}$& ID$^{b}$& $\log P$ & $<V >$ & $<J>$ & $<H>$ & $<K_S>$ & $A_V $ & $A_J$ &  $A_H$ & $A_K$ &JD$_{mean}^V$ & Sample\\  
  &  & [days] & [mag] & [mag] & [mag] & [mag] &[mag] & [mag] & [mag] &[mag] &[HJD] & \\ 
   \hline 
  \endhead 
    \hline
  \multicolumn{13}{l}{$^{(a)}$ Primary identification.}\\
 \multicolumn{13}{l}{$^{(b)}$ OGLE identification for LMC and SMC Cepheids.} \\
    \hline
\endfoot
  \hline
  \multicolumn{13}{l}{$^{(a)}$ Primary identification.}\\
 \multicolumn{13}{l}{$^{(b)}$ OGLE identification for LMC and SMC Cepheids.} \\
    \hline
\endlastfoot

    &  &  &  &  & &  && & & && \\
\multicolumn{13}{c}{MW} \\
AQ	Pup	&&	1.47865	&	8.987	&	6.029	&	5.481	&	5.288	&	1.23	&	0.57	&	0.51	&	0.50	&	2444161.277	&	LS\\
$\beta$	Dor	&&	0.99311	&	3.752	&	2.383	&	2.029	&	1.944	&	0.62	&	0.26	&	0.21	&	0.20	&	2452359.566	&	LS\\
CV	Mon	&&	0.73067	&	6.962	&	7.407	&	$\cdots$	&	$\cdots$	&	0.22	&	0.29	&	$\cdots$	&	$\cdots$	&	2445935.188	&	LS\\
FM	Aql	&&	0.78634	&	8.272	&	5.704	&	5.219	&	5.040	&	0.74	&	0.31	&	0.21	&	0.20	&	2431317.201	&	LS\\
KN	Cen	&&	1.53192	&	10.015	&	6.438	&	5.755	&	5.485	&	1.22	&	0.48	&	0.43	&	0.42	&	2452438.273	&	LS\\
KQ	Sco	&&	1.45777	&	9.81	&	5.944	&	5.216	&	4.943	&	0.89	&	0.36	&	0.37	&	0.38	&	2444058.473	&	LS\\
RZ	Gem	&&	0.74265	&	10.019	&	7.645	&	7.159	&	6.957	&	0.97	&	0.38	&	0.27	&	0.16	&	2445058.781	&	LS\\
S	Nor	&&	0.98919	&	6.427	&	4.67	&	4.274	&	4.148	&	0.66	&	0.28	&	0.20	&	0.20	&	2444035.766	&	LS\\
S	Sge	&&	0.92335	&	5.618	&	4.178	&	3.845	&	3.750	&	0.71	&	0.34	&	0.25	&	0.24	&	2442694.531	&	LS\\
ST	Tau	&&	0.60577	&	8.22	&	6.327	&	5.923	&	5.795	&	0.77	&	0.31	&	0.23	&	0.21	&	2444924.312	&	LS\\
SZ	Aql	&&	1.23401	&	8.614	&	5.885	&	5.346	&	5.150	&	1.24	&	0.44	&	0.40	&	0.40	&	2445314.453	&	LS\\
T	Mon	&&	1.43178	&	5.51	&	4.122	&	3.653	&	3.513	&	1.01	&	0.47	&	0.46	&	0.43	&	2444296.535	&	LS\\
TT	Aql	&&	1.13846	&	7.128	&	4.683	&	4.186	&	4.015	&	1.06	&	0.4	&	0.37	&	0.36	&	2431801.854	&	LS\\
T	Vul	&&	0.64693	&	5.751	&	4.554	&	4.237	&	4.164	&	0.68	&	0.27	&	0.14	&	0.14	&	2441655.809	&	LS\\
U	Aql	&&	0.84659	&	3.987	&	4.397	&	3.987	&	3.839	&	0.23	&	0.31	&	0.23	&	0.23	&	2443315.785	&	LS\\
U	Car	&&	1.58891	&	6.281	&	4.13	&	3.668	&	3.509	&	1.18	&	0.51	&	0.46	&	0.45	&	2444608.883	&	LS\\
U	Nor	&&	1.10189	&	9.234	&	5.857	&	5.237	&	4.984	&	0.97	&	0.38	&	0.25	&	0.24	&	2449565.789	&	LS\\
U	Sgr	&&	0.829	&	6.692	&	4.525	&	4.092	&	3.942	&	0.71	&	0.29	&	0.22	&	0.20	&	2443128.703	&	LS\\
UU	Mus	&&	1.06581	&	9.779	&	7.466	&	6.990	&	6.818	&	1.09	&	0.42	&	0.31	&	0.29	&	2434857.191	&	LS\\
V	Cen	&&	0.73988	&	6.823	&	5.074	&	$\cdots$	&	$\cdots$	&	0.77	&	0.31	&	$\cdots$	&	$\cdots$	&	2452372.520	&	LS\\
V496	Aql	&&	0.83296	&	7.745	&	5.561	&	5.124	&	4.965	&	0.37	&	0.16	&	0.12	&	0.12	&	2445014.871	&	LS\\
VW	Cen	&&	1.17717	&	10.244	&	7.586	&	7.015	&	6.811	&	1.02	&	0.38	&	0.34	&	0.33	&	2452383.797	&	LS\\
VY	Car	&&	1.27662	&	7.458	&	5.412	&	4.944	&	4.793	&	1.1	&	0.46	&	0.38	&	0.38	&	2450928.727	&	LS\\
W	Sgr	&&	0.88053	&	4.669	&	3.239	&	2.909	&	2.811	&	0.78	&	0.31	&	0.26	&	0.24	&	2443844.566	&	LS\\
WZ	Sgr	&&	1.33944	&	8.026	&	5.333	&	4.763	&	4.558	&	1.11	&	0.44	&	0.41	&	0.40	&	2441884.863	&	LS\\
X	Cyg	&&	1.21447	&	6.399	&	4.41	&	3.953	&	3.816	&	0.98	&	0.39	&	0.37	&	0.37	&	2443861.48	&	LS\\
XX	Cen	&&	1.03955	&	7.817	&	5.933	&	5.531	&	5.394	&	0.91	&	0.4	&	0.25	&	0.25	&	2452379.812	&	LS\\
Y	Oph	&&	1.23366	&	6.151	&	3.375	&	2.873	&	2.677	&	0.49	&	0.19	&	0.14	&	0.13	&	2442727.805	&	LS\\
Y	Sgr	&&	0.76143	&	5.744	&	4.062	&	3.703	&	3.572	&	0.74	&	0.28	&	0.22	&	0.21	&	2443648.270	&	LS\\
$\zeta$	Gem	&&	1.00646	&	3.913	&	2.556	&	2.200	&	2.115	&	0.48	&	0.21	&	0.17	&	0.17	&	2452369.785	&	LS\\
AA	Gem	&&	1.0532	&	9.717	&	7.648	&	7.191	&	7.086	&	0.73	&	0.22	&	0.22	&	0.23	&	2454508.062	&	MP\\
AA	Mon	&&	0.5953	&	12.595	&	9.702	&	$\cdots$	&	$\cdots$	&	0.74	&	0.31	&	$\cdots$	&	$\cdots$	&	2454500.935	&	MP\\
AA	Ser	&&	1.234	&	12.235	&	7.544	&	6.758	&	6.492	&	0.97	&	0.34	&	0.33	&	0.28	&	2454577.305	&	MP\\
AC	Mon	&&	0.9039	&	10.084	&	7.573	&	7.070	&	$\cdots$	&	0.73	&	0.33	&	0.26	&	$\cdots$	&	2454765.882	&	MP\\
AD	Gem	&&	0.5784	&	9.864	&	8.466	&	8.141	&	$\cdots$	&	0.69	&	0.27	&	0.20	&	$\cdots$	&	2454592.444	&	MP\\
AK	Cep	&&	0.8593	&	11.193	&	8.403	&	7.892	&	7.744	&	0.67	&	0.24	&	0.21	&	0.21	&	2454495.813	&	MP\\
AN	Aur	&&	1.01238	&	10.451	&	7.936	&	7.427	&	7.266	&	0.74	&	0.33	&	0.25	&	0.20	&	2436861.067	&	MP\\
AO	Aur	&&	0.8301	&	10.851	&	8.642	&	8.181	&	8.059	&	0.95	&	0.37	&	0.28	&	0.25	&	2454500.251	&	MP\\
AS	Vul	&&	1.0872	&	12.25	&	8.377	&	$\cdots$	&	7.491	&	0.95	&	0.34	&	$\cdots$	&	0.30	&	2454611.727	&	MP\\
AV	Tau	&&	0.5582	&	12.326	&	9.194	&	8.636	&	8.443	&	0.89	&	0.36	&	0.27	&	0.26	&	2454523.719	&	MP\\
AW	Per	&&	0.81047	&	7.466	&	5.224	&	4.819	&	4.694	&	0.84	&	0.34	&	0.30	&	0.23	&	2450115.616	&	MP\\
AX	Aur	&&	0.4838	&	12.442	&	$\cdots$	&	$\cdots$	&	9.322	&	0.59	&	$\cdots$	&	$\cdots$	&	0.11	&	2454770.153	&	MP\\
AY	Sgr	&&	0.8175	&	10.483	&	7.123	&	6.519	&	6.299	&	0.99	&	0.33	&	0.23	&	0.23	&	2454655.283	&	MP\\
BG	Lac	&&	0.72688	&	8.879	&	7.029	&	6.646	&	6.524	&	0.7	&	0.27	&	0.20	&	0.19	&	2441552.792	&	MP\\
BK	Aur	&&	0.9032	&	9.473	&	$\cdots$	&	$\cdots$	&	6.752	&	0.64	&	$\cdots$	&	$\cdots$	&	0.21	&	2454779.649	&	MP\\
BM	Per	&&	1.36094	&	10.405	&	6.658	&	5.992	&	5.741	&	1.32	&	0.65	&	0.57	&	0.52	&	2447007.365	&	MP\\
BR	Vul	&&	0.7158	&	10.694	&	7.717	&	7.201	&	7.026	&	0.83	&	0.35	&	0.25	&	0.23	&	2454631.057	&	MP\\
BV	Mon	&&	0.4792	&	11.396	&	$\cdots$	&	$\cdots$	&	8.447	&	0.6	&	$\cdots$	&	$\cdots$	&	0.22	&	2454522.900	&	MP\\
BZ	Cyg	&&	1.00611	&	10.221	&	6.750	&	6.134	&	5.894	&	0.55	&	0.27	&	0.20	&	0.17	&	2443791.486	&	MP\\
CD	Cas	&&	0.8921	&	10.786	&	7.630	&	7.072	&	6.890	&	1.68	&	0.31	&	0.24	&	0.22	&	2454772.401	&	MP\\
CD	Cyg	&&	1.23234	&	8.964	&	6.353	&	5.840	&	5.686	&	1.28	&	0.46	&	0.40	&	0.41	&	2443864.507	&	MP\\
CF	Cas	&&	0.68798	&	11.133	&	8.600	&	8.125	&	7.945	&	0.65	&	0.28	&	0.22	&	0.19	&	2444153.728	&	MP\\
CH	Cas	&&	1.1786	&	10.988	&	7.325	&	6.677	&	6.413	&	1.15	&	0.39	&	0.29	&	0.29	&	2454765.623	&	MP\\
CK	Sct	&&	0.87017	&	10.593	&	$\cdots$	&	$\cdots$	&	6.629	&	0.5	&	$\cdots$	&	$\cdots$	&	0.18	&	2444295.112	&	MP\\
CN	Cep	&&	0.9778	&	12.351	&	8.317	&	$\cdots$	&	7.355	&	0.65	&	0.27	&	$\cdots$	&	0.15	&	2454756.562	&	MP\\
CN	Sct	&&	0.9997	&	12.481	&	7.821	&	7.059	&	6.734	&	0.68	&	0.28	&	0.23	&	0.26	&	2454678.161	&	MP\\
CP	Cep	&&	1.25196	&	10.577	&	7.331	&	6.710	&	6.511	&	0.86	&	0.34	&	0.35	&	0.33	&	2433084.259	&	MP\\
CR	Cep	&&	0.79471	&	9.626	&	$\cdots$	&	6.079	&	5.903	&	0.37	&	$\cdots$	&	0.15	&	0.13	&	2445285.259	&	MP\\
CR	Ser	&&	0.7244	&	10.843	&	$\cdots$	&	6.747	&	6.520	&	0.74	&	$\cdots$	&	0.22	&	0.19	&	2454747.903	&	MP\\
CS	Ori	&&	0.5899	&	11.361	&	9.345	&	$\cdots$	&	8.832	&	0.95	&	0.37	&	$\cdots$	&	0.27	&	2454588.501	&	MP\\
CY	Aur	&&	1.1414	&	11.866	&	8.577	&	$\cdots$	&	7.737	&	0.92	&	0.37	&	$\cdots$	&	0.31	&	2454793.174	&	MP\\
CY	Cas	&&	1.1577	&	11.647	&	7.845	&	$\cdots$	&	6.935	&	1.2	&	0.45	&	$\cdots$	&	0.42	&	2454661.217	&	MP\\
DD	Cas	&&	0.99174	&	9.878	&	7.533	&	7.053	&	6.920	&	0.69	&	0.22	&	0.21	&	0.20	&	2444249.993	&	MP\\
DL	Cas	&&	0.90311	&	8.963	&	6.554	&	6.088	&	5.930	&	0.66	&	0.24	&	0.22	&	0.17	&	2444963.195	&	MP\\
EP	Cyg	&&	0.6323	&	12.744	&	$\cdots$	&	$\cdots$	&	9.529	&	0.84	&	$\cdots$	&	$\cdots$	&	0.24	&	2454524.868	&	MP\\
ER	Aur	&&	1.1956	&	11.525	&	9.069	&	8.593	&	8.449	&	0.65	&	0.22	&	0.16	&	0.13	&	2454582.214	&	MP\\
EZ	Cyg	&&	1.0667	&	11.049	&	$\cdots$	&	7.636	&	7.478	&	0.9	&	$\cdots$	&	0.30	&	0.28	&	2454681.277	&	MP\\
FM	Cas	&&	0.76412	&	9.12	&	$\cdots$	&	6.752	&	6.619	&	0.58	&	$\cdots$	&	0.19	&	0.22	&	2444234.299	&	MP\\
FN	Aql	&&	0.97688	&	8.384	&	5.965	&	5.482	&	5.334	&	0.73	&	0.30	&	0.21	&	0.23	&	2444074.928	&	MP\\
GH	Cyg	&&	0.8931	&	9.886	&	$\cdots$	&	$\cdots$	&	6.616	&	0.75	&	$\cdots$	&	$\cdots$	&	0.19	&	2454782.548	&	MP\\
GQ	Vul	&&	1.1019	&	13.612	&	$\cdots$	&	7.933	&	7.664	&	1.04	&	$\cdots$	&	0.31	&	0.30	&	2454756.430	&	MP\\
GX	SGE	&&	1.1106	&	12.464	&	8.261	&	$\cdots$	&	$\cdots$	&	1.11	&	0.37	&	$\cdots$	&	$\cdots$	&	2454685.276	&	MP\\
GY	SGE	&&	1.71347	&	10.156	&	5.552	&	4.848	&	4.562	&	0.86	&	0.32	&	0.26	&	0.21	&	2447528.080	&	MP\\
HZ	Per	&&	1.0523	&	13.774	&	9.126	&	8.322	&	8.035	&	0.84	&	0.33	&	0.35	&	0.31	&	2454764.436	&	MP\\
KX	Cyg	&&	1.302	&	11.948	&	6.834	&	$\cdots$	&	5.622	&	1.17	&	0.42	&	$\cdots$	&	0.37	&	2454681.859	&	MP\\
MM	Per	&&	0.6147	&	10.803	&	8.664	&	8.253	&	8.127	&	0.66	&	0.29	&	0.20	&	0.21	&	2454764.489	&	MP\\
MW	Cyg	&&	0.77486	&	9.484	&	$\cdots$	&	6.197	&	6.017	&	0.74	&	$\cdots$	&	0.23	&	0.22	&	2444572.574	&	MP\\
OT	Per	&&	1.4165	&	13.479	&	8.720	&	7.875	&	7.601	&	0.99	&	0.44	&	0.44	&	0.41	&	2454811.329	&	MP\\
RR	Lac	&&	0.80729	&	8.848	&	6.988	&	6.614	&	6.506	&	0.87	&	0.30	&	0.23	&	0.24	&	2444123.329	&	MP\\
RS	Cas	&&	0.79906	&	9.945	&	6.768	&	6.213	&	6.005	&	0.92	&	0.30	&	0.23	&	0.25	&	2442785.239	&	MP\\
RT	Aur	&&	0.57151	&	5.468	&	4.251	&	3.982	&	3.923	&	0.84	&	0.29	&	0.17	&	0.19	&	2434106.322	&	MP\\
RU	Sct	&&	1.29454	&	9.473	&	5.888	&	5.285	&	5.054	&	1.16	&	0.41	&	0.40	&	0.36	&	2443270.680	&	MP\\
RW	Cam	&&	1.21527	&	8.652	&	5.820	&	5.278	&	5.112	&	0.93	&	0.37	&	0.35	&	0.34	&	2444988.157	&	MP\\
RW	Cas	&&	1.1701	&	9.238	&	6.836	&	6.355	&	6.211	&	1.39	&	0.43	&	0.37	&	0.37	&	2454510.707	&	MP\\
RX	Aur	&&	1.06536	&	7.673	&	5.743	&	5.345	&	5.250	&	0.8	&	0.30	&	0.26	&	0.20	&	2443758.430	&	MP\\
RX	Cam	&&	0.89829	&	7.671	&	$\cdots$	&	$\cdots$	&	4.580	&	0.72	&	$\cdots$	&	$\cdots$	&	0.24	&	2444933.522	&	MP\\
RY	Cas	&&	1.08415	&	9.9	&	$\cdots$	&	6.537	&	6.347	&	0.95	&	$\cdots$	&	0.24	&	0.28	&	2437366.744	&	MP\\
RY	Cma	&&	0.6701	&	8.107	&	6.386	&	6.028	&	5.911	&	0.8	&	0.30	&	0.26	&	0.23	&	2454746.464	&	MP\\
RZ	Cma	&&	0.6289	&	9.697	&	7.551	&	7.177	&	7.013	&	0.66	&	0.28	&	0.24	&	0.20	&	2454498.298	&	MP\\
SS	Sct	&&	0.56482	&	8.215	&	6.306	&	5.925	&	5.825	&	0.6	&	0.23	&	0.16	&	0.13	&	2443105.589	&	MP\\
SU	Cyg	&&	0.58496	&	6.867	&	5.673	&	5.383	&	5.320	&	0.8	&	0.41	&	0.22	&	0.22	&	2443309.086	&	MP\\
SV	Mon	&&	1.18283	&	8.258	&	6.265	&	5.822	&	5.708	&	1.21	&	0.41	&	0.38	&	0.42	&	2445057.698	&	MP\\
SV	Per	&&	1.04647	&	8.972	&	$\cdots$	&	6.345	&	6.214	&	0.81	&	$\cdots$	&	0.28	&	0.26	&	2445017.787	&	MP\\
S	Vul	&&	1.83703	&	8.972	&	$\cdots$	&	4.817	&	$\cdots$	&	0.55	&	$\cdots$	&	0.20	&	$\cdots$	&	2446397.463	&	MP\\
SV	Vul	&&	1.65348	&	7.23	&	$\cdots$	&	4.073	&	3.917	&	0.99	&	$\cdots$	&	0.28	&	0.25	&	2441955.528	&	MP\\
SW	Cas	&&	0.73567	&	9.696	&	7.413	&	6.980	&	6.847	&	0.8	&	0.28	&	0.22	&	0.19	&	2445002.037	&	MP\\
SX	Per	&&	0.6325	&	11.15	&	8.768	&	$\cdots$	&	8.189	&	0.81	&	0.37	&	$\cdots$	&	0.25	&	2454770.118	&	MP\\
SY	Aur	&&	1.0062	&	9.077	&	6.924	&	6.524	&	$\cdots$	&	0.69	&	0.28	&	0.20	&	$\cdots$	&	2454537.304	&	MP\\
SY	Cas	&&	0.60971	&	9.886	&	7.831	&	7.446	&	7.333	&	0.93	&	0.32	&	0.23	&	0.20	&	2441689.917	&	MP\\
SZ	Cyg	&&	1.17926	&	9.429	&	6.516	&	5.945	&	5.747	&	1.02	&	0.35	&	0.34	&	0.31	&	2444952.43	&	MP\\
TV	Cma	&&	0.6693	&	10.6	&	8.031	&	7.557	&	7.391	&	0.76	&	0.34	&	0.24	&	0.22	&	2454774.442	&	MP\\
TW	Cma	&&	0.8448	&	9.558	&	$\cdots$	&	7.170	&	7.049	&	0.63	&	$\cdots$	&	0.24	&	0.25	&	2454748.029	&	MP\\
TX	Cyg	&&	1.1676	&	9.492	&	5.313	&	$\cdots$	&	4.340	&	1.26	&	0.39	&	$\cdots$	&	0.28	&	2454671.006	&	MP\\
TX	Mon	&&	0.9396	&	10.955	&	8.583	&	$\cdots$	&	7.941	&	0.66	&	0.31	&	$\cdots$	&	0.21	&	2454758.498	&	MP\\
TY	Mon	&&	0.6045	&	11.722	&	$\cdots$	&	$\cdots$	&	8.707	&	0.62	&	$\cdots$	&	$\cdots$	&	0.20	&	2454499.999	&	MP\\
TY	Sct	&&	1.04351	&	10.812	&	7.225	&	6.621	&	6.403	&	0.96	&	0.36	&	0.25	&	0.26	&	2444968.673	&	MP\\
U	Vul	&&	0.90259	&	7.123	&	4.526	&	4.074	&	3.931	&	0.81	&	0.30	&	0.21	&	0.20	&	2443268.403	&	MP\\
UZ	Sct	&&	1.16872	&	11.234	&	$\cdots$	&	6.735	&	6.496	&	0.86	&	$\cdots$	&	0.28	&	0.26	&	2444196.693	&	MP\\
V1162	Aql	&&	0.7305	&	7.807	&	$\cdots$	&	5.805	&	5.703	&	0.52	&	$\cdots$	&	0.18	&	0.17	&	2454677.842	&	MP\\
V1344	Aql	&&	0.8738	&	7.793	&	5.181	&	$\cdots$	&	4.572	&	0.41	&	0.16	&	$\cdots$	&	0.11	&	2454740.306	&	MP\\
V1364	Cyg	&&	1.1132	&	13.269	&	$\cdots$	&	$\cdots$	&	8.082	&	0.7	&	$\cdots$	&	$\cdots$	&	0.23	&	2454788.747	&	MP\\
V1467	Cyg	&&	1.6866	&	13.48	&	8.150	&	7.279	&	6.961	&	1.09	&	0.47	&	0.41	&	0.39	&	2454850.753	&	MP\\
V	Lac	&&	0.6975	&	8.883	&	$\cdots$	&	$\cdots$	&	6.570	&	1.87	&	$\cdots$	&	$\cdots$	&	0.22	&	2454763.571	&	MP\\
VV	Cas	&&	0.7929	&	10.747	&	8.323	&	7.855	&	7.725	&	1.01	&	0.34	&	0.21	&	0.21	&	2454760.472	&	MP\\
VX	Cyg	&&	1.30391	&	10.056	&	$\cdots$	&	6.065	&	5.875	&	1.0	&	$\cdots$	&	0.39	&	0.36	&	2443822.170	&	MP\\
VX	Per	&&	1.03688	&	9.282	&	$\cdots$	&	6.423	&	$\cdots$	&	0.58	&	$\cdots$	&	0.22	&	$\cdots$	&	2445045.821	&	MP\\
VY	Cyg	&&	0.89527	&	9.561	&	$\cdots$	&	6.536	&	6.373	&	0.82	&	$\cdots$	&	0.27	&	0.25	&	2444474.496	&	MP\\
VZ	Cyg	&&	0.68703	&	8.959	&	$\cdots$	&	6.840	&	4.609	&	0.67	&	$\cdots$	&	0.21	&	1.13	&	2441714.810	&	MP\\
W	Gem	&&	0.8984	&	6.941	&	$\cdots$	&	$\cdots$	&	4.689	&	0.75	&	$\cdots$	&	$\cdots$	&	0.20	&	2454506.588	&	MP\\
WW	Mon	&&	0.6686	&	12.519	&	$\cdots$	&	$\cdots$	&	9.387	&	1.07	&	$\cdots$	&	$\cdots$	&	0.32	&	2454524.68	&	MP\\
X	Sct	&&	0.623	&	9.962	&	$\cdots$	&	$\cdots$	&	6.800	&	0.52	&	$\cdots$	&	$\cdots$	&	0.17	&	2454670.380	&	MP\\
X	Vul	&&	0.80069	&	8.841	&	5.898	&	5.393	&	5.229	&	0.81	&	0.31	&	0.20	&	0.22	&	2444478.853	&	MP\\
XY	Cas	&&	0.6534	&	9.999	&	$\cdots$	&	$\cdots$	&	7.197	&	0.47	&	$\cdots$	&	$\cdots$	&	0.20	&	2454746.639	&	MP\\
Y	Aur	&&	0.5865	&	9.659	&	7.662	&	7.276	&	7.145	&	0.87	&	0.35	&	0.25	&	0.21	&	2454489.710	&	MP\\
Y	Lac	&&	0.63586	&	9.153	&	7.637	&	7.299	&	7.216	&	0.79	&	0.32	&	0.19	&	0.20	&	2441754.858	&	MP\\
Y	Sct	&&	1.01457	&	9.6	&	$\cdots$	&	5.878	&	5.661	&	0.78	&	$\cdots$	&	0.26	&	0.26	&	2447432.248	&	MP\\
YZ	Aur	&&	1.2599	&	10.333	&	$\cdots$	&	6.889	&	6.705	&	0.84	&	$\cdots$	&	0.37	&	0.39	&	2437465.107	&	MP\\
YZ	Sgr	&&	0.98017	&	7.347	&	5.386	&	4.986	&	4.881	&	0.82	&	0.31	&	0.22	&	0.21	&	2443125.835	&	MP\\
Z	Lac	&&	1.03686	&	8.416	&	6.241	&	5.800	&	5.670	&	1.08	&	0.41	&	0.30	&	0.28	&	2442846.860	&	MP\\
Z	Sct	&&	1.1106	&	9.587	&	$\cdots$	&	$\cdots$	&	6.300	&	0.89	&	$\cdots$	&	$\cdots$	&	0.27	&	2454686.814	&	MP\\
   &  &  &  &  & &  && & & && \\
\multicolumn{13}{c}{LMC} \\
HV1003	&	OGLE-LMC-CEP-2504 &	1.38625	&	13.183	&	$\cdots$	&	$\cdots$	&	11.394	&	0.9	&	$\cdots$	&	$\cdots$	&	0.36	&	2453041.092	&	P04\\
HV1005	&	OGLE-LMC-CEP-2534 &	1.27221	&	14.084	&	12.698	&	$\cdots$	&	$\cdots$	&	1.11	&	0.55	&	$\cdots$	&	$\cdots$	&	2452215.58	&	P04\\
HV1006	&	OGLE-LMC-CEP-2636 &	1.15279	&	14.413	&	12.931	&	12.377	&	12.268	&	1.22	&	0.42	&	0.37	&	0.36	&	2452201.675	&	P04\\
HV1019	&	OGLE-LMC-CEP-3013 &	1.13543	&	14.363	&	13.040	&	12.480	&	12.376	&	0.71	&	0.27	&	0.25	&	0.26	&	2450752.191	&	P04\\
HV1023	&	OGLE-LMC-CEP-3158 &	1.42423	&	13.81	&	$\cdots$	&	11.616	&	11.499	&	1.1	&	$\cdots$	&	0.44	&	0.43	&	2453053.341	&	P04\\
HV12471	&	OGLE-LMC-CEP-0174 &	1.20038	&	14.759	&	$\cdots$	&	12.457	&	12.310	&	0.78	&	$\cdots$	&	0.32	&	0.31	&	2452179.952	&	P04\\
HV12656	&	OGLE-LMC-CEP-3203 &	1.12711	&	14.152	&	$\cdots$	&	$\cdots$	&	12.373	&	0.59	&	$\cdots$	&	$\cdots$	&	0.13	&	2452209.891	&	P04\\
HV12724	&	OGLE-LMC-CEP-0070 &	1.13817	&	14.761	&	13.246	&	$\cdots$	&	12.543	&	0.8	&	0.32	&	$\cdots$	&	0.35	&	2452201.152	&	P04\\
HV2244	&	OGLE-LMC-CEP-0500 &	1.14545	&	14.215	&	12.956	&	12.420	&	$\cdots$	&	1.06	&	0.35	&	0.33	&	$\cdots$	&	2452184.736	&	P04\\
HV2251	&	OGLE-LMC-CEP-0654 &	1.44556	&	13.262	&	11.841	&	11.327	&	11.224	&	1.35	&	0.60	&	0.54	&	0.51	&	2453047.383	&	P04\\
HV2257	&	OGLE-LMC-CEP-0512 &	1.59547	&	13.13	&	11.506	&	$\cdots$	&	10.844	&	1.19	&	0.57	&	$\cdots$	&	0.47	&	2453063.08	&	P04\\
HV2260	&	OGLE-LMC-CEP-0655 &	1.11353	&	14.883	&	13.430	&	12.855	&	12.733	&	0.93	&	0.41	&	0.40	&	0.37	&	2452190.166	&	P04\\
HV2270	&	OGLE-LMC-CEP-0648 &	1.13437	&	14.651	&	13.204	&	12.622	&	12.517	&	0.83	&	0.36	&	0.33	&	0.33	&	2450748.565	&	P04\\
HV2282	&	OGLE-LMC-CEP-0683 &	1.16664	&	14.362	&	12.948	&	$\cdots$	&	12.301	&	1.04	&	0.38	&	$\cdots$	&	0.38	&	2452185.867	&	P04\\
HV2324	&	OGLE-LMC-CEP-1001 &	1.16038	&	14.361	&	12.977	&	12.416	&	12.308	&	0.93	&	0.37	&	0.33	&	0.33	&	2450747.540	&	P04\\
HV2339	&	OGLE-LMC-CEP-1088 &	1.14236	&	14.319	&	$\cdots$	&	12.370	&	$\cdots$	&	0.75	&	$\cdots$	&	0.26	&	$\cdots$	&	2450739.811	&	P04\\
HV2369	&	OGLE-LMC-CEP-1290 &	1.68465	&	12.671	&	11.080	&	10.532	&	10.413	&	1.17	&	0.49	&	0.43	&	0.42	&	2453121.998	&	P04\\
HV2432	&	OGLE-LMC-CEP-1538 &	1.03815	&	14.226	&	13.154	&	12.655	&	12.571	&	0.63	&	0.37	&	0.29	&	0.25	&	2452141.289	&	P04\\
HV2463	&	OGLE-LMC-CEP-1641 &	1.14496	&	14.124	&	12.904	&	12.382	&	12.297	&	1.08	&	0.45	&	0.29	&	0.28	&	2452134.698	&	P04\\
HV2527	&	OGLE-LMC-CEP-1954 &	1.11226	&	14.627	&	13.244	&	12.682	&	12.570	&	1.08	&	0.37	&	0.34	&	0.34	&	2452189.202	&	P04\\
HV2538	&	OGLE-LMC-CEP-2030 &	1.14212	&	14.472	&	13.036	&	12.466	&	12.356	&	0.53	&	0.28	&	0.25	&	0.27	&	2452192.164	&	P04\\
HV2793	&	OGLE-LMC-CEP-2949 &	1.2838	&	14.103	&	12.556	&	$\cdots$	&	$\cdots$	&	1.0	&	0.37	&	$\cdots$	&	$\cdots$	&	2452220.997	&	P04\\
HV5655	&	OGLE-LMC-CEP-1184 &	1.15264	&	14.55	&	13.139	&	12.552	&	$\cdots$	&	0.93	&	0.40	&	0.33	&	$\cdots$	&	2452189.321	&	P04\\
HV8036	&	OGLE-LMC-CEP-0068 &	1.45304	&	13.568	&	$\cdots$	&	11.557	&	$\cdots$	&	1.09	&	$\cdots$	&	0.44	&	$\cdots$	&	2453065.819	&	P04\\
HV873	&	OGLE-LMC-CEP-0328 &	1.53731	&	13.023	&	11.614	&	11.098	&	10.989	&	1.27	&	0.59	&	0.49	&	0.45	&	2453065.424	&	P04\\
HV875	&	OGLE-LMC-CEP-0434 &	1.48206	&	13.043	&	11.767	&	11.285	&	11.193	&	0.64	&	0.30	&	0.23	&	0.18	&	2453049.035	&	P04\\
HV876	&	OGLE-LMC-CEP-0467 &	1.35636	&	13.65	&	12.297	&	11.770	&	11.668	&	1.07	&	0.58	&	0.51	&	0.50	&	2452189.214	&	P04\\
HV878	&	OGLE-LMC-CEP-0501 &	1.36749	&	13.541	&	12.249	&	11.738	&	11.631	&	1.21	&	0.56	&	0.50	&	0.50	&	2452211.414	&	P04\\
HV881	&	OGLE-LMC-CEP-0528 &	1.55282	&	13.104	&	11.672	&	11.146	&	$\cdots$	&	1.25	&	0.58	&	0.44	&	$\cdots$	&	2453062.365	&	P04\\
HV882	&	OGLE-LMC-CEP-0590 &	1.50231	&	13.383	&	11.860	&	11.334	&	$\cdots$	&	1.23	&	0.45	&	0.46	&	$\cdots$	&	2453053.958	&	P04\\
HV885	&	OGLE-LMC-CEP-0712 &	1.31606	&	13.627	&	$\cdots$	&	11.623	&	11.501	&	0.93	&	$\cdots$	&	0.22	&	0.24	&	2450746.376	&	P04\\
HV887	&	OGLE-LMC-CEP-0727 &	1.16104	&	14.101	&	12.843	&	12.325	&	12.245	&	1.1	&	0.41	&	0.35	&	0.35	&	2450746.860	&	P04\\
HV889	&	OGLE-LMC-CEP-0821 &	1.41167	&	13.746	&	12.183	&	11.626	&	11.511	&	1.02	&	0.46	&	0.42	&	0.40	&	2450770.531	&	P04\\
HV892	&	OGLE-LMC-CEP-0848 &	1.20383	&	14.22	&	$\cdots$	&	12.343	&	$\cdots$	&	1.08	&	$\cdots$	&	0.42	&	$\cdots$	&	2450754.57	&	P04\\
HV893	&	OGLE-LMC-CEP-0935 &	1.32465	&	13.883	&	12.446	&	11.877	&	11.769	&	0.93	&	0.52	&	0.46	&	0.45	&	2450761.350	&	P04\\
HV899	&	OGLE-LMC-CEP-0986 &	1.49207	&	13.366	&	11.911	&	11.362	&	11.249	&	1.21	&	0.54	&	0.48	&	0.48	&	2450779.712	&	P04\\
HV900	&	OGLE-LMC-CEP-0966 &	1.67564	&	12.76	&	11.304	&	10.746	&	10.626	&	0.92	&	0.45	&	0.39	&	0.39	&	2453075.211	&	P04\\
HV911	&	OGLE-LMC-CEP-1166 &	1.14329	&	14.347	&	13.047	&	12.502	&	$\cdots$	&	1.05	&	0.38	&	0.38	&	$\cdots$	&	2450474.096	&	P04\\
HV932	&	OGLE-LMC-CEP-1578 &	1.12331	&	14.176	&	12.986	&	$\cdots$	&	12.387	&	1.1	&	0.44	&	$\cdots$	&	0.33	&	2450474.456	&	P04\\
HV997	&	OGLE-LMC-CEP-2337 &	1.11869	&	14.597	&	13.146	&	12.574	&	12.454	&	0.97	&	0.41	&	0.32	&	0.25	&	2452184.360	&	P04\\
U1	&	OGLE-LMC-CEP-0079 &	1.35302	&	14.115	&	12.509	&	$\cdots$	&	$\cdots$	&	1.11	&	0.53	&	$\cdots$	&	$\cdots$	&	2452190.014	&	P04\\
   &  &  &  &  & &  && & & && \\
\multicolumn{13}{c}{SMC} \\
  &	OGLE-SMC-CEP-1107 	&	1.38884	&	14.687	&	12.841	&	12.377	&	12.291	&	0.93	&	0.36	&	0.44	&	0.40	&	2450648.891	&	SMC\\
&	OGLE-SMC-CEP-1108 	&	0.76017	&	15.69	&	14.465	&	$\cdots$	&	$\cdots$	&	0.68	&	0.24	&	$\cdots$	&	$\cdots$	&	2450629.508	&	SMC\\
&	OGLE-SMC-CEP-1130 	&	0.56735	&	16.097	&	$\cdots$	&	14.754	&	14.732	&	1.14	&	$\cdots$	&	0.28	&	0.21	&	2450627.434	&	SMC\\
&	OGLE-SMC-CEP-1163 	&	0.70814	&	16.648	&	$\cdots$	&	14.593	&	14.507	&	1.01	&	$\cdots$	&	0.27	&	0.24	&	2450627.969	&	SMC\\
&	OGLE-SMC-CEP-1172 	&	1.14809	&	14.897	&	$\cdots$	&	12.909	&	12.846	&	0.88	&	$\cdots$	&	0.27	&	0.27	&	2450639.828	&	SMC\\
&	OGLE-SMC-CEP-1195 	&	0.5681	&	16.062	&	14.980	&	$\cdots$	&	$\cdots$	&	1.23	&	0.41	&	$\cdots$	&	$\cdots$	&	2450470.504	&	SMC\\
&	OGLE-SMC-CEP-1205 	&	0.70875	&	16.217	&	14.867	&	$\cdots$	&	$\cdots$	&	0.99	&	0.30	&	$\cdots$	&	$\cdots$	&	2450473.473	&	SMC\\
&	OGLE-SMC-CEP-1247 	&	1.1239	&	15.334	&	$\cdots$	&	13.246	&	13.121	&	0.82	&	$\cdots$	&	0.30	&	0.26	&	2450492.000	&	SMC\\
&	OGLE-SMC-CEP-1261 	&	1.0017	&	15.318	&	$\cdots$	&	13.498	&	13.440	&	0.56	&	$\cdots$	&	0.20	&	0.20	&	2450476.129	&	SMC\\
&	OGLE-SMC-CEP-1331 	&	0.46872	&	16.527	&	15.440	&	$\cdots$	&	$\cdots$	&	1.22	&	0.42	&	$\cdots$	&	$\cdots$	&	2450469.684	&	SMC\\
&	OGLE-SMC-CEP-1340 	&	1.2438	&	15.144	&	$\cdots$	&	12.863	&	12.797	&	0.86	&	$\cdots$	&	0.32	&	0.34	&	2450493.102	&	SMC\\
&	OGLE-SMC-CEP-1363 	&	0.68537	&	15.887	&	$\cdots$	&	14.320	&	14.299	&	0.95	&	$\cdots$	&	0.27	&	0.26	&	2450472.043	&	SMC\\
&	OGLE-SMC-CEP-1377 	&	1.12918	&	14.425	&	13.189	&	$\cdots$	&	$\cdots$	&	0.85	&	0.23	&	$\cdots$	&	$\cdots$	&	2452103.086	&	SMC\\
&	OGLE-SMC-CEP-1385 	&	1.19926	&	14.767	&	13.259	&	12.790	&	12.668	&	0.97	&	0.28	&	0.24	&	0.24	&	2450494.457	&	SMC\\
&	OGLE-SMC-CEP-1403 	&	1.4591	&	14.358	&	12.639	&	12.132	&	12.016	&	0.56	&	0.26	&	0.26	&	0.25	&	2450494.602	&	SMC\\
&	OGLE-SMC-CEP-1410 	&	0.96322	&	15.217	&	13.851	&	13.425	&	13.321	&	0.92	&	0.28	&	0.26	&	0.24	&	2450479.812	&	SMC\\
&	OGLE-SMC-CEP-1438 	&	0.98044	&	15.021	&	$\cdots$	&	13.419	&	13.403	&	0.92	&	$\cdots$	&	0.23	&	0.19	&	2450478.227	&	SMC\\
&	OGLE-SMC-CEP-1453 	&	0.57515	&	16.898	&	$\cdots$	&	14.868	&	14.757	&	0.92	&	$\cdots$	&	0.27	&	0.27	&	2450472.844	&	SMC\\
&	OGLE-SMC-CEP-1477 	&	1.50546	&	13.957	&	12.327	&	11.813	&	11.750	&	1.15	&	0.42	&	0.39	&	0.42	&	2450522.379	&	SMC\\
&	OGLE-SMC-CEP-1481 	&	0.85513	&	16.096	&	$\cdots$	&	14.122	&	14.006	&	0.7	&	$\cdots$	&	0.24	&	0.21	&	2450477.359	&	SMC\\
&	OGLE-SMC-CEP-1492 	&	0.79877	&	15.214	&	14.154	&	13.809	&	13.728	&	1.03	&	0.30	&	0.24	&	0.20	&	2452094.098	&	SMC\\
&	OGLE-SMC-CEP-1524 	&	1.43794	&	14.182	&	12.514	&	12.038	&	11.963	&	0.95	&	0.36	&	0.38	&	0.38	&	2450496.633	&	SMC\\
&	OGLE-SMC-CEP-1538 	&	0.76367	&	15.491	&	$\cdots$	&	14.032	&	13.895	&	1.04	&	$\cdots$	&	0.28	&	0.21	&	2452095.309	&	SMC\\
&	OGLE-SMC-CEP-1549 	&	0.83365	&	15.492	&	14.236	&	$\cdots$	&	$\cdots$	&	0.97	&	0.32	&	$\cdots$	&	$\cdots$	&	2450474.098	&	SMC\\
&	OGLE-SMC-CEP-1569 	&	1.19447	&	14.58	&	13.166	&	12.736	&	12.669	&	0.94	&	0.24	&	0.26	&	0.25	&	2450481.770	&	SMC\\
&	OGLE-SMC-CEP-1632 	&	0.65325	&	16.048	&	$\cdots$	&	14.364	&	14.371	&	1.02	&	$\cdots$	&	0.25	&	0.22	&	2450473.094	&	SMC\\
&	OGLE-SMC-CEP-1635 	&	1.34539	&	14.386	&	$\cdots$	&	12.333	&	12.270	&	1.0	&	$\cdots$	&	0.44	&	0.40	&	2450491.879	&	SMC\\
&	OGLE-SMC-CEP-1681 	&	0.85904	&	15.498	&	14.184	&	13.818	&	13.764	&	0.99	&	0.27	&	0.23	&	0.24	&	2450473.430	&	SMC\\
&	OGLE-SMC-CEP-1686 	&	1.54016	&	13.561	&	12.048	&	$\cdots$	&	$\cdots$	&	0.48	&	0.14	&	$\cdots$	&	$\cdots$	&	2450671.691	&	SMC\\
&	OGLE-SMC-CEP-1693 	&	0.89775	&	15.632	&	$\cdots$	&	13.745	&	13.673	&	0.63	&	$\cdots$	&	0.20	&	0.23	&	2450481.199	&	SMC\\
&	OGLE-SMC-CEP-1723 	&	1.21579	&	14.99	&	$\cdots$	&	12.896	&	12.779	&	0.8	&	$\cdots$	&	0.30	&	0.32	&	2450497.105	&	SMC\\
&	OGLE-SMC-CEP-1747 	&	1.28572	&	13.955	&	12.683	&	12.290	&	12.198	&	1.03	&	0.30	&	0.31	&	0.29	&	2450501.504	&	SMC\\
&	OGLE-SMC-CEP-1797 	&	1.61542	&	13.783	&	$\cdots$	&	11.529	&	11.443	&	0.58	&	$\cdots$	&	0.23	&	0.23	&	2450534.965	&	SMC\\
&	OGLE-SMC-CEP-1857 	&	0.69391	&	15.569	&	14.518	&	$\cdots$	&	$\cdots$	&	1.1	&	0.34	&	$\cdots$	&	$\cdots$	&	2452094.477	&	SMC\\
&	OGLE-SMC-CEP-1913 	&	1.19061	&	14.705	&	13.215	&	12.756	&	12.686	&	0.86	&	0.26	&	0.31	&	0.29	&	2450492.816	&	SMC\\
&	OGLE-SMC-CEP-1960 	&	0.61994	&	16.018	&	$\cdots$	&	14.503	&	14.439	&	0.96	&	$\cdots$	&	0.25	&	0.21	&	2450472.848	&	SMC\\
&	OGLE-SMC-CEP-1984 	&	0.9788	&	14.972	&	13.656	&	13.297	&	13.225	&	0.83	&	0.19	&	0.23	&	0.22	&	2452103.223	&	SMC\\
&	OGLE-SMC-CEP-2024 	&	0.63712	&	15.368	&	14.380	&	14.050	&	13.954	&	1.1	&	0.35	&	0.25	&	0.20	&	2450473.523	&	SMC\\
&	OGLE-SMC-CEP-2031 	&	1.16354	&	14.523	&	13.248	&	12.890	&	12.831	&	0.8	&	0.27	&	0.27	&	0.25	&	2450490.043	&	SMC\\
&	OGLE-SMC-CEP-2049 	&	0.46313	&	16.021	&	15.077	&	$\cdots$	&	$\cdots$	&	1.14	&	0.37	&	$\cdots$	&	$\cdots$	&	2452090.727	&	SMC\\
&	OGLE-SMC-CEP-2060 	&	0.59858	&	16.145	&	15.027	&	$\cdots$	&	$\cdots$	&	0.81	&	0.39	&	$\cdots$	&	$\cdots$	&	2450471.223	&	SMC\\
&	OGLE-SMC-CEP-2066 	&	0.30169	&	16.743	&	15.825	&	$\cdots$	&	$\cdots$	&	1.28	&	0.47	&	$\cdots$	&	$\cdots$	&	2450470.074	&	SMC\\
&	OGLE-SMC-CEP-2068 	&	0.74747	&	16.091	&	14.791	&	$\cdots$	&	$\cdots$	&	0.85	&	0.32	&	$\cdots$	&	$\cdots$	&	2450475.680	&	SMC\\
&	OGLE-SMC-CEP-2090 	&	0.92075	&	15.261	&	13.960	&	13.573	&	13.518	&	0.87	&	0.27	&	0.25	&	0.21	&	2450482.145	&	SMC\\
&	OGLE-SMC-CEP-2134 	&	1.19346	&	14.372	&	13.045	&	$\cdots$	&	$\cdots$	&	0.87	&	0.19	&	$\cdots$	&	$\cdots$	&	2452107.660	&	SMC\\
&	OGLE-SMC-CEP-2143 	&	0.87495	&	15.563	&	$\cdots$	&	13.919	&	13.819	&	0.89	&	$\cdots$	&	0.27	&	0.25	&	2450473.902	&	SMC\\
&	OGLE-SMC-CEP-2149 	&	0.66106	&	15.719	&	14.663	&	14.342	&	14.260	&	0.94	&	0.31	&	0.20	&	0.16	&	2450471.016	&	SMC\\
&	OGLE-SMC-CEP-2202 	&	0.95664	&	14.958	&	$\cdots$	&	13.341	&	13.319	&	0.84	&	$\cdots$	&	0.25	&	0.22	&	2450479.879	&	SMC\\
&	OGLE-SMC-CEP-2230 	&	1.06606	&	14.659	&	13.418	&	13.038	&	12.991	&	0.99	&	0.32	&	0.21	&	0.19	&	2452103.043	&	SMC\\
&	OGLE-SMC-CEP-2266 	&	0.45931	&	15.974	&	15.031	&	$\cdots$	&	$\cdots$	&	1.11	&	0.42	&	$\cdots$	&	$\cdots$	&	2450625.074	&	SMC\\
&	OGLE-SMC-CEP-2269 	&	1.50494	&	13.721	&	$\cdots$	&	11.750	&	11.655	&	0.33	&	$\cdots$	&	0.15	&	0.15	&	2450675.227	&	SMC\\
&	OGLE-SMC-CEP-2280 	&	0.69167	&	15.799	&	14.652	&	$\cdots$	&	$\cdots$	&	0.89	&	0.28	&	$\cdots$	&	$\cdots$	&	2450632.617	&	SMC\\
&	OGLE-SMC-CEP-2291 	&	0.36786	&	16.928	&	15.835	&	$\cdots$	&	$\cdots$	&	1.08	&	0.39	&	$\cdots$	&	$\cdots$	&	2450624.973	&	SMC\\
&	OGLE-SMC-CEP-2313 	&	1.05685	&	14.797	&	13.496	&	$\cdots$	&	$\cdots$	&	0.9	&	0.25	&	$\cdots$	&	$\cdots$	&	2450641.695	&	SMC\\
&	OGLE-SMC-CEP-2318 	&	0.71607	&	15.589	&	14.492	&	14.139	&	14.053	&	1.01	&	0.30	&	0.28	&	0.25	&	2450629.637	&	SMC\\
&	OGLE-SMC-CEP-2326 	&	0.59883	&	16.186	&	14.903	&	14.533	&	14.487	&	1.02	&	0.32	&	0.29	&	0.26	&	2450625.527	&	SMC\\
&	OGLE-SMC-CEP-2329 	&	1.51492	&	13.978	&	12.283	&	11.851	&	11.715	&	0.91	&	0.31	&	0.36	&	0.34	&	2450671.496	&	SMC\\
&	OGLE-SMC-CEP-2376 	&	0.41226	&	16.165	&	15.292	&	$\cdots$	&	$\cdots$	&	1.03	&	0.38	&	$\cdots$	&	$\cdots$	&	2450625.465	&	SMC\\
&	OGLE-SMC-CEP-2384 	&	0.72624	&	15.694	&	14.530	&	$\cdots$	&	$\cdots$	&	1.04	&	0.30	&	$\cdots$	&	$\cdots$	&	2450630.508	&	SMC\\
&	OGLE-SMC-CEP-2451 	&	0.62864	&	15.835	&	14.620	&	$\cdots$	&	$\cdots$	&	0.74	&	0.22	&	$\cdots$	&	$\cdots$	&	2450627.270	&	SMC\\
&	OGLE-SMC-CEP-2454 	&	0.8905	&	15.159	&	$\cdots$	&	13.591	&	13.549	&	0.98	&	$\cdots$	&	0.24	&	0.23	&	2452095.523	&	SMC\\
&	OGLE-SMC-CEP-2456 	&	0.54651	&	16.206	&	$\cdots$	&	14.781	&	14.763	&	1.17	&	$\cdots$	&	0.33	&	0.28	&	2450626.062	&	SMC\\
&	OGLE-SMC-CEP-2538 	&	0.68641	&	16.223	&	14.834	&	$\cdots$	&	$\cdots$	&	0.73	&	0.27	&	$\cdots$	&	$\cdots$	&	2450630.070	&	SMC\\
&	OGLE-SMC-CEP-2577 	&	0.59491	&	15.94	&	14.783	&	$\cdots$	&	$\cdots$	&	0.91	&	0.25	&	$\cdots$	&	$\cdots$	&	2450625.684	&	SMC\\
&	OGLE-SMC-CEP-2606 	&	0.57704	&	16.076	&	$\cdots$	&	14.624	&	14.544	&	1.06	&	$\cdots$	&	0.26	&	0.22	&	2450625.617	&	SMC\\
&	OGLE-SMC-CEP-2607 	&	1.0317	&	15.238	&	$\cdots$	&	13.285	&	13.223	&	0.48	&	$\cdots$	&	0.17	&	0.18	&	2450636.684	&	SMC\\
&	OGLE-SMC-CEP-2634 	&	0.39962	&	16.784	&	15.739	&	$\cdots$	&	$\cdots$	&	1.09	&	0.42	&	$\cdots$	&	$\cdots$	&	2452091.031	&	SMC\\
&	OGLE-SMC-CEP-2700 	&	0.36563	&	16.995	&	15.870	&	$\cdots$	&	$\cdots$	&	1.3	&	0.42	&	$\cdots$	&	$\cdots$	&	2450626.066	&	SMC\\
&	OGLE-SMC-CEP-2707 	&	0.78316	&	16.009	&	14.729	&	14.340	&	14.234	&	0.98	&	0.32	&	0.21	&	0.25	&	2450631.902	&	SMC\\
&	OGLE-SMC-CEP-2712 	&	0.69008	&	15.557	&	14.444	&	$\cdots$	&	$\cdots$	&	1.02	&	0.30	&	$\cdots$	&	$\cdots$	&	2450629.121	&	SMC\\
&	OGLE-SMC-CEP-2721 	&	0.64466	&	15.77	&	14.701	&	$\cdots$	&	$\cdots$	&	1.1	&	0.33	&	$\cdots$	&	$\cdots$	&	2450628.734	&	SMC\\
&	OGLE-SMC-CEP-2722 	&	0.69785	&	15.787	&	14.605	&	14.202	&	14.088	&	1.04	&	0.30	&	0.22	&	0.15	&	2450630.078	&	SMC\\
&	OGLE-SMC-CEP-2747 	&	0.65913	&	15.981	&	$\cdots$	&	14.462	&	14.396	&	1.07	&	$\cdots$	&	0.25	&	0.30	&	2450630.430	&	SMC\\
&	OGLE-SMC-CEP-2824 	&	0.87506	&	15.279	&	14.114	&	13.755	&	13.692	&	0.88	&	0.29	&	0.23	&	0.23	&	2450633.055	&	SMC\\
&	OGLE-SMC-CEP-2835 	&	0.89958	&	15.484	&	14.169	&	$\cdots$	&	$\cdots$	&	0.51	&	0.19	&	$\cdots$	&	$\cdots$	&	2450636.758	&	SMC\\
&	OGLE-SMC-CEP-2836 	&	0.32379	&	17.059	&	$\cdots$	&	15.621	&	$\cdots$	&	1.27	&	$\cdots$	&	0.38	&	$\cdots$	&	2450624.754	&	SMC\\
&	OGLE-SMC-CEP-2837 	&	0.6324	&	16.138	&	14.907	&	$\cdots$	&	$\cdots$	&	1.15	&	0.38	&	$\cdots$	&	$\cdots$	&	2450628.301	&	SMC\\
&	OGLE-SMC-CEP-2841 	&	1.1677	&	14.968	&	13.346	&	12.906	&	12.859	&	0.92	&	0.27	&	0.26	&	0.29	&	2450644.539	&	SMC\\
&	OGLE-SMC-CEP-2877 	&	0.48521	&	16.235	&	15.176	&	$\cdots$	&	$\cdots$	&	0.76	&	0.28	&	$\cdots$	&	$\cdots$	&	2450624.875	&	SMC\\
&	OGLE-SMC-CEP-2887 	&	0.83314	&	15.685	&	$\cdots$	&	14.019	&	13.982	&	0.89	&	$\cdots$	&	0.26	&	0.25	&	2450634.309	&	SMC\\
&	OGLE-SMC-CEP-2890 	&	0.574	&	16.097	&	14.999	&	$\cdots$	&	$\cdots$	&	1.11	&	0.35	&	$\cdots$	&	$\cdots$	&	2450625.895	&	SMC\\
&	OGLE-SMC-CEP-2905 	&	1.58032	&	13.994	&	12.213	&	11.782	&	11.558	&	0.53	&	0.18	&	0.23	&	0.23	&	2450686.488	&	SMC\\
&	OGLE-SMC-CEP-2909 	&	0.50706	&	16.414	&	15.325	&	$\cdots$	&	$\cdots$	&	1.21	&	0.35	&	$\cdots$	&	$\cdots$	&	2450626.168	&	SMC\\
&	OGLE-SMC-CEP-2920 	&	0.59802	&	16.145	&	15.032	&	14.647	&	14.624	&	1.17	&	0.31	&	0.29	&	0.24	&	2450628.129	&	SMC\\
&	OGLE-SMC-CEP-2933 	&	0.39695	&	16.468	&	15.482	&	$\cdots$	&	$\cdots$	&	1.26	&	0.45	&	$\cdots$	&	$\cdots$	&	2452091.098	&	SMC\\
&	OGLE-SMC-CEP-2947 	&	0.93866	&	15.306	&	14.060	&	13.638	&	13.556	&	0.71	&	0.26	&	0.22	&	0.19	&	2450632.281	&	SMC\\
&	OGLE-SMC-CEP-2968 	&	0.948	&	15.019	&	13.765	&	13.384	&	13.319	&	0.84	&	0.22	&	0.21	&	0.21	&	2450638.449	&	SMC\\
&	OGLE-SMC-CEP-2971 	&	0.33465	&	16.52	&	15.585	&	$\cdots$	&	$\cdots$	&	1.21	&	0.41	&	$\cdots$	&	$\cdots$	&	2450625.711	&	SMC\\
&	OGLE-SMC-CEP-2977 	&	0.57227	&	16.544	&	$\cdots$	&	14.803	&	14.771	&	0.74	&	$\cdots$	&	0.24	&	0.16	&	2450630.574	&	SMC\\
&	OGLE-SMC-CEP-3003 	&	0.59266	&	15.85	&	14.847	&	14.505	&	14.450	&	1.02	&	0.30	&	0.32	&	0.29	&	2450629.105	&	SMC\\
&	OGLE-SMC-CEP-3020 	&	0.56421	&	15.778	&	14.714	&	$\cdots$	&	$\cdots$	&	1.19	&	0.30	&	$\cdots$	&	$\cdots$	&	2450626.000	&	SMC\\
&	OGLE-SMC-CEP-3045 	&	0.90041	&	14.915	&	$\cdots$	&	13.512	&	13.453	&	0.91	&	$\cdots$	&	0.20	&	0.16	&	2450632.996	&	SMC\\
&	OGLE-SMC-CEP-3046 	&	0.66619	&	15.789	&	14.766	&	14.411	&	14.340	&	1.01	&	0.33	&	0.20	&	0.19	&	2452094.652	&	SMC\\
&	OGLE-SMC-CEP-3056 	&	0.49577	&	16.353	&	15.276	&	14.974	&	14.931	&	1.21	&	0.35	&	0.29	&	0.25	&	2450626.262	&	SMC\\
&	OGLE-SMC-CEP-3062 	&	0.73052	&	15.841	&	$\cdots$	&	14.211	&	14.101	&	0.85	&	$\cdots$	&	0.25	&	0.17	&	2450630.023	&	SMC\\
&	OGLE-SMC-CEP-3074 	&	0.73709	&	16.046	&	14.795	&	14.404	&	14.345	&	0.57	&	0.21	&	0.19	&	0.17	&	2450628.469	&	SMC\\
&	OGLE-SMC-CEP-3095 	&	0.47766	&	16.103	&	$\cdots$	&	14.858	&	14.856	&	1.11	&	$\cdots$	&	0.26	&	0.22	&	2452090.230	&	SMC\\
&	OGLE-SMC-CEP-3112 	&	0.63253	&	15.899	&	$\cdots$	&	14.425	&	14.415	&	0.87	&	$\cdots$	&	0.27	&	0.27	&	2450625.848	&	SMC\\
&	OGLE-SMC-CEP-3139 	&	1.21071	&	14.641	&	13.128	&	12.661	&	12.545	&	0.86	&	0.26	&	0.30	&	0.27	&	2450645.141	&	SMC\\
&	OGLE-SMC-CEP-3163 	&	0.47322	&	16.051	&	15.073	&	14.735	&	14.697	&	1.09	&	0.40	&	0.27	&	0.24	&	2450626.582	&	SMC\\
&	OGLE-SMC-CEP-3227 	&	0.41652	&	16.638	&	15.564	&	$\cdots$	&	$\cdots$	&	1.3	&	0.35	&	$\cdots$	&	$\cdots$	&	2452090.633	&	SMC\\
&	OGLE-SMC-CEP-3284 	&	0.81232	&	15.814	&	14.584	&	14.240	&	14.160	&	1.01	&	0.30	&	0.30	&	0.23	&	2452098.703	&	SMC\\
&	OGLE-SMC-CEP-3302 	&	0.33092	&	16.785	&	$\cdots$	&	15.465	&	$\cdots$	&	1.16	&	$\cdots$	&	0.32	&	$\cdots$	&	2452089.723	&	SMC\\
&	OGLE-SMC-CEP-3305 	&	1.11192	&	14.837	&	13.495	&	13.048	&	12.994	&	0.82	&	0.27	&	0.33	&	0.27	&	2450644.312	&	SMC\\

     \hline

\end{longtable}
\end{landscape}

%
\clearpage

\begin{table}
\scriptsize
\caption{Pulsation parameters for FO calibrating Cepheids }
\label{tab_cat_fo}
\centering
\begin{tabular}{llrrrrc}
\hline\hline

OGLE ID&$\log P$ &<V> &<J> &$A_V $&$A_J $& JD$_{mean}^V$\\
 &
[days]&
 [mag]&
[mag]&
  [mag]&
 [mag]&
 [HJD]\\
\hline

 OGLE-SMC-CEP-2002   & 0.40123  & 15.785  & 14.805  & 0.48  & 0.19  &       2452089.804\\
OGLE-SMC-CEP-2043   & 0.54709  & 15.790  & 14.650  & 0.43  & 0.15  &        2450472.057\\
OGLE-SMC-CEP-2371   & 0.37723  & 16.597  & 15.520  & 0.47  & 0.20  &        2450625.333\\
OGLE-SMC-CEP-2816   & 0.18456  & 16.464  & 15.629  & 0.53  & 0.22  &        2450623.134\\
OGLE-SMC-CEP-2948   & 0.45681  & 15.942  & 14.911  & 0.47  & 0.13  &        2452127.708\\
OGLE-SMC-CEP-3040   & 0.39686  & 15.863  & 14.812  & 0.46  & 0.18  &        2450629.533\\
OGLE-SMC-CEP-3082   & 0.22626  & 16.322  & 15.431  & 0.48  & 0.20  &        2452088.542\\
OGLE-SMC-CEP-3126   & 0.15992  & 16.416  & 15.568  & 0.61  & 0.23  &        2450623.298\\
OGLE-SMC-CEP-3183   & 0.29055  & 16.329  & 15.388  & 0.50  & 0.24  &        2452088.954\\
OGLE-SMC-CEP-3298   & 0.42815  & 15.677  & 14.675  & 0.43  & 0.16  &        2450623.992\\

\hline
\end{tabular}
\end{table}

%
\clearpage
\onecolumn

\begin{table}
\scriptsize
\caption{F7 and G3 coefficients of the  FU NIR light-curve templates and F3 and G2 coefficients of the FO $J$-band templates}
\label{tab_c}
\centering
\begin{tabular}{rrrrrrrrrrrrrrrrrc}

\hline\hline

 BIN & A$_0$ &
 A$_1$ &
 A$_2$ &
 A$_3$ &
 A$_4$ &
 A$_5$ &
 A$_6$ &
 A$_7$ &
 $\Phi_1$ &
 $\Phi_2$ &
 $\Phi_3$ &
 $\Phi_4$ &
 $\Phi_5$ &
 $\Phi_6$ &
 $\Phi_7$ &
 $\sigma_F$
 \\
&&&&&&&&&&&&&&&&&\\
\hline
\multicolumn{18}{c}{$J$-band} \\
  1&  -0.002 & 0.325 & 0.174 & 0.107 & 0.076 & 0.046 & 0.032 & 0.022 & 1.583 & 1.610 &1.648 & 1.633& 1.543 & 1.412 & 0.995&0.05\\
  2&  -0.001 & 0.373 & 0.172 & 0.092  &  0.051 &   0.032  &  0.017 &   0.012 & 1.460  &    1.652   &   1.702   &   1.728  &    1.814  &    1.657    &  1.661 &0.05\\
  3&   0.003 & 0.418 & 0.156 & 0.077 &  0.028&    0.014 &  0.003  & 0.007 &1.423   &   1.823 &     1.903    & 1.843    &  1.310    & 0.951   & 1.921 &0.04\\
  4&  0.004 &  0.430 &    0.134  &  0.083  &  0.016  &  0.012  &  0.012 &   0.005  & 1.355 &  2.011   &   2.114   &   1.261    &  2.292   &   1.135    &  1.849 & 0.04\\
  5&  -0.018 & 0.424 & 0.040 & 0.002 & 0.028 &0.019 &0.004 & 0.015 & 1.494 & -3.112 & 0.199 & 0.404 & 2.548 & 0.643 & 1.396 & 0.04\\
  6&  0.000 & 0.412 & 0.076 & 0.028 & 0.021 & 0.011 & 0.011 & 0.005 & 1.616 & 2.225 & 0.787  & 0.811 & 0.952 & -0.018 &  1.117 &0.05\\
  7&  -0.002 & 0.486 & 0.093 & 0.063 & 0.063 & 0.045 & 0.045 & 0.024 &1.421 & 2.070  & 1.581  & 1.535  & 1.591  & 1.910 & 2.081 & 0.05\\
  8&  -0.002 & 0.477 & 0.106 & 0.067 & 0.060 & 0.046 & 0.017 & 0.015 & 1.416 & 2.026 & 1.623 & 1.550 & 1.776 & 2.188 & 2.045 & 0.04\\
  9&   -0.003 & 0.424  & 0.114 &  0.067 &  0.039 &  0.030 &  0.012 &  0.012 & 1.218  & 1.731  & 2.331 &  2.652 &  -3.129 &  -2.609  & -2.159 & 0.03\\
  10&  -0.004 & 0.413 &  0.133 &  0.067 &  0.047 &  0.033 &  0.016 &  0.015 & 1.215  & 1.802 &  2.115 &  2.409 &  2.390 &  2.912  & -2.382   & 0.04\\

        FO &0.001&0.454&0.056&0.019&  $\cdots$ &  $\cdots$ &  $\cdots$ &  $\cdots$ 
   &1.586&1.586&3.099&  $\cdots$ &  $\cdots$ &  $\cdots$ &  $\cdots$ &0.07\\
     &&&&&&&&&&&&&&&&\\
\multicolumn{18}{c}{$H$-band} \\
  1&  -0.013  &  0.386  & 0.135 &  0.057 &  0.014  & 0.009  &  0.004 &  0.003  & 1.338  & 2.020 &  2.381  & 2.423  & -3.111  & 2.000  & 1.658 & 0.05\\
  2&   -0.012 &  0.388  & 0.135  & 0.057 &  0.014 &  0.008 & 0.003 &  0.002 & 1.332  & 2.017  & 2.406  & 2.435  & -3.096 &  1.943  & 1.614 & 0.05\\
  3&  -0.003 &  0.432  & 0.120 &  0.066  & 0.020  & 0.007  & 0.005  & 0.006 &  1.291  & 2.104  & 2.702 &  2.437 &  2.389  & -2.527  & 2.398 & 0.04\\
  4&  -0.003 &  0.424  & 0.111 &  0.066 &  0.009 & 0.004 & 0.002 &  0.003 & 1.253  & 2.241 &  2.900 &  2.695  & -3.061  & 1.498 &  3.054 & 0.04\\
  5&  0.003 &  0.462  & 0.069 &  0.016  & 0.027 &  0.023  & 0.017 &  0.012 & 1.456 &  3.058 &  0.011 &  2.419 &  -1.323 &  1.461  & 3.091 & 0.05\\
  6&  -0.002 &  0.480  & 0.072  & 0.019 &  0.017 &  0.006 &  0.006  & 0.004  &  1.396  & 2.878 &  2.012  & 2.754  & 2.726  & 1.223 &  1.922 & 0.03\\
  7&  0.000 & 0.485  & 0.056  & 0.031 &  0.013 &  0.010 &  0.006 &  0.004 &  1.415 &  2.667 &  2.528 &  2.478  & -2.969 &  -2.867  & -1.826 & 0.03\\
  8&  -0.001 & 0.476  & 0.064 &  0.039  & 0.013 &  0.007 &  0.010  & 0.005 &  1.369  & 2.498 &  2.920 &  3.009  & -2.705  & -1.954 &  -0.790 & 0.03\\
  9&  -0.003 & 0.451  & 0.094  & 0.046 &  0.023 &  0.013 & 0.008 & 0.001 & 1.330  & 2.328  & -3.079  & -2.651  & -1.692  & -0.741  & 1.184 & 0.03\\
  10&  -0.006  & 0.451  & 0.092  & 0.035 &  0.011  & 0.006  & 0.003 &  0.001 & 1.314 &  2.505  & -2.720 &  -1.883 &  -0.768  & 1.090  & -2.898 & 0.03\\
       &&&&&&&&&&&&&&&&\\
\multicolumn{18}{c}{$K_{\rm{S}}$-band} \\
  1 &  -0.002 & 0.367 &   0.119  &  0.042 &   0.008  &  0.019 &   0.014 &   0.011  &   1.382  &  1.734  &  2.923  &  -3.029  &  2.088 &   2.537 &   2.960 &  0.07\\
  2 &  -0.001 & 0.450  &  0.119  & 0.058  & 0.025  & 0.020  & 0.018  & 0.005  & 1.325  & 1.954  & 2.831 &  -2.976  & 2.638 & 2.960  & -2.407 & 0.05\\
  3 & 0.001 & 0.458  & 0.121 &  0.047&  0.018 &  0.016  & 0.001 & 0.007  & 1.313  & 2.269  & 2.813  & 2.750  & -2.949 &  -2.027  & 2.853 & 0.05\\
  4 & 0.000 & 0.423 &  0.099 &  0.064 &  0.016 &  0.001 &  0.008  & 0.005 & 1.230  & 2.410  & -2.945 &  -3.124  & 0.201 &  2.615  & -2.382 & 0.04\\
  5 & 0.003 & 0.423 &  0.099 &  0.064 &  0.016 &  0.001 &  0.008  & 0.005  & 1.230  & 2.410  & -2.945 &  -3.124  & 0.201 &  2.615  & -2.382 & 0.05\\
  6 & 0.004 & 0.488 & 0.078  & 0.024 &  0.017 & 0.016 &  0.004 & 0.005  & 1.375 &  2.994  & 2.442 &  2.527  &  2.639  &  1.917 &  3.100 & 0.04\\
  7 &  0.002 & 0.483  & 0.054 &  0.035 &  0.018 & 0.016 &  0.012 &  0.006 & 1.409  & 2.536  & 2.890  & -3.028  & -2.626 &  -1.715  & 0.689 & 0.04\\
  8 &  0.005 & 0.481  & 0.066  & 0.031 &  0.009  & 0.006 &  0.008 &  0.002  & 1.386  & 2.687 &  2.997  & 2.973  & -2.037 &  -1.625  & 0.457 & 0.03\\
  9 &-0.001 &  0.473 &   0.076 &   0.032  &  0.021 &   0.014 &   0.010  &  0.011 &   1.382  &  2.370  &  3.052 &   -2.547 &   -2.207 &   -1.495 &   -0.092 &  0.03\\
  10 & -0.001  &  0.460  &  0.098 &   0.039 &   0.021 &   0.010  &   0.010  &  0.003 &   1.311  &  2.403  &  -3.088  &  -2.344  &  -1.865 &   -0.718  &  0.999 & 0.03\\
   &&&&&&&&&& &&&&&&\\

\hline
\hline
 BIN &
 G$_1$ &
 G$_2$ &
 G$_3$ &
 $\Gamma_1$ &
 $\Gamma_2$ &
 $\Gamma_3$ &
 $\tau_1$ &
 $\tau_2$ &
 $\tau_3$ &
 &
&
&
&
&
&
 $\sigma_G$\\
 &&&&&&&&&& &&&&&&\\
\hline

\multicolumn{18}{c}{$J$--band} \\
     1& 1.213 & -1.285 & 0.629 & 0.880 & -0.026 & 0.945 & 0.660 & 0.708 & 0.226  &&&&&&& 0.06\\
     2& 0.577 & -0.728 & 0.800 & 0.934 & -0.003 & 0.822 & 0.269 & 0.871 & 0.631  &&&&&&& 0.05\\
     3& 2.227 & -1.882 & -0.536 & 0.976 & 0.076 & 0.054 & 0.972 & 1.096 & -0.344  &&&&&&& 0.04\\
     4& 0.633 & -0.418 & 0.427 & 0.770 & 0.157 & 0.901 & 0.650 & 1.448 & 0.295  &&&&&&& 0.05\\
     5& 0.672 & -0.489 & 0.113 & 0.748 & 0.265 & 0.920 & 0.720 & 1.144 & -0.347  &&&&&&& 0.05\\
     6& 0.674 & -0.451 & 0.450 & 0.732 & 0.102 & 0.933 & 0.558 & 1.124 & 0.352  &&&&&&& 0.04\\
     7& 1.320 & -0.835 & 0.438 & 0.774 & -0.108 & 0.942 & 0.786 & 1.899 & -0.176  &&&&&&& 0.04\\
     8& 0.476 & -0.785 & 1.233 & 0.938 & -0.097 & 0.781 & 0.200 & 1.497 & 0.717  &&&&&&& 0.04\\
     9& 0.823 & -0.606 & -0.384 & 0.905 & 0.234 & 0.033 & 0.777 & 0.919 & 0.281  &&&&&&& 0.04\\
     10& 0.655 & -0.339 & 0.315 & 0.880 & 0.253 & 0.718 & 0.376 & 1.125 & -0.472  &&&&&&& 0.04\\

          FO & 1.002&-1.023& $\cdots$ &0.863 & 0.144 &$\cdots$ &1.095 & 1.055 &$\cdots$ &&&&&&& 0.07\\
                & &  &  &  &  &  &  &  &&&&&&&\\
\multicolumn{18}{c}{$H$--band} \\
     1&0.837 & -0.485 & 0.422 & 0.870 & -0.089 & 0.685 & 0.443 & 1.466 & -0.573  &&&&&&& 0.05\\ 
     2&0.931 & -0.513 & -0.387 & 0.887 & 0.192 & 0.027 & 0.728 & 1.169 & 0.381  &&&&&&& 0.05\\
     3&0.395 & -0.357 & 0.509 & 0.744 & 0.272 & 0.871 & 0.511 & 1.003 & -0.348  &&&&&&& 0.04\\
     4&0.455 & -0.374 & 0.336 & 0.767 & 0.286 & 0.876 & 0.461 & 0.770 & 0.304  &&&&&&& 0.04\\
     5&0.128 & -0.471 & 0.716 & 0.647 & 0.251 & 0.790 & 0.211 & 1.117 & 0.662  &&&&&&& 0.06\\
     6&0.857 & -0.578 & 0.234 & 0.722 & 0.370 & 0.892 & 0.730 & 1.448 & 0.343  &&&&&&& 0.03\\
     7&1.051 & -0.888 & -0.203 & 0.805 & 0.294 & 0.033 & 1.094 & 1.333 & 0.340  &&&&&&& 0.03\\
     8&0.374 & -0.447 & 0.457 & 0.866 & 0.293 & 0.724 & 0.323 & 0.768 & 0.549  &&&&&&& 0.03\\
     9&0.284 & -0.419 & 0.506 & 0.858 & 0.288 & 0.766 & 0.281 & 0.827 & 0.565  &&&&&&& 0.03\\
     10&0.605 & -0.411 & 0.223 & 0.836 & 0.287 & 0.696 & 0.391 & 0.736 & -0.396  &&&&&&& 0.03\\
         & &  &  &  &  &  &  &  &&&&&&&\\
\multicolumn{18}{c}{$K_{\rm{S}}$--band} \\
     1&0.552 & -0.375 & 0.139 & 0.811 & 0.219 & 0.916 & 0.472 & 0.692 & 0.164  &&&&&&& 0.07\\
     2&0.552 & -0.375 & 0.139 & 0.811 & 0.219 & 0.916 & 0.472 & 0.692 & 0.164  &&&&&&& 0.04\\
     3&0.403 & -0.450 & 0.695 & 0.642 & 0.397 & 0.852 & 0.673 & 1.429 & 0.415  &&&&&&& 0.05\\
     4&0.494 & -0.384 & 0.276 & 0.767 & 0.297 & 0.878 & 0.421 & 0.691 & 0.290  &&&&&&& 0.04\\
     5&0.634 & -0.464 & 0.246 & 0.722 & 0.299 & 0.875 & 0.659 & 1.247 & 0.526  &&&&&&& 0.05\\
     6&0.737 & -0.496 & 0.417 & 0.826 & 0.384 & 0.629 & 0.568 & 1.556 & 0.542  &&&&&&& 0.04\\
     7&0.539 & -0.542 & 0.333 & 0.699 & 0.322 & 0.849 & 0.763 & 0.988 & 0.377  &&&&&&& 0.04\\
     8&0.599 & -0.477 & 0.359 & 0.832 & 0.334 & 0.634 & 0.476 & 1.005 & 0.592  &&&&&&& 0.03\\
     9&0.648 & -0.499 & 0.184 & 0.783 & 0.262 & 0.849 & 0.703 & 1.037 & -0.256  &&&&&&& 0.03\\
     10&0.525 & -0.442 & 0.195 & 0.779 & 0.278 & 0.849 & 0.547 & 0.797 & 0.375  &&&&&&& 0.03\\

\hline
\end{tabular}
\end{table}

%
\clearpage
\onecolumn
\scriptsize

\begin{table}
\caption{NIR--to--optical amplitude ratios.}
\label{tab_amp}
\centering
\begin{tabular}{l|ccc|ccc}
\hline\hline
~&\multicolumn{3}{c}{\textbf{MW+ LMC} } &\multicolumn{3}{|c}{\textbf{SMC}} \\
 \hline
  Period   &   $A_J$/$A_V$ & $A_H$/$A_V$ & $A_{Ks}$/$A_V$ & $A_J$/$A_V$ & $A_H$/$A_V$ &
   $A_{Ks}$/$A_V$ \\
\hline
 P $\le$20 days$^{a}$          &0.43 $\pm$ 0.01$^{b}$ & 0.34 $\pm$ 0.01$^{b}$ & 0.33$\pm$ 0.01 $^{b}$ &0.33  $\pm$ 0.01 $^{b}$ &0.26 $\pm$0.01$^{b}$&0.25$\pm$0.01$^{b}$\\
 P $>$20 days$^{a}$            &0.45 $\pm$ 0.01$^{b}$ &0.40 $\pm$ 0.01$^{b}$ &0.41 $\pm$ 0.01$^{b}$ &0.39 $\pm$0.03$^{b}$&0.39$\pm$0.03$^{b}$&0.37$\pm$0.02$^{b}$\\
\hline
\multicolumn{7}{l}{$^{(a)}$Note that for the $A_{Ks}$/$A_V$ ratio of SMC Cepheids the adopted period ranges are P$\le$15.5 days and P$>$15.5 days.} \\
\multicolumn{7}{l}{$^{(b)}$ These errors were estimated as 
$rms/\sqrt{N}$, where $N$ is the number of calibrators in the specific period range.}
\end{tabular}
\end{table}

\clearpage

\end{document}